\documentclass[oldversion,preprint]{aa}
\usepackage{graphicx}
\usepackage{txfonts}
\begin{document}

\title{A decade of SN\,1993J: discovery of radio wavelength effects in the expansion rate}

\authorrunning{J.M. Marcaide et al.}
\titlerunning{A decade of expansion of SN\,1993J}

\author{J.M. Marcaide\inst{1} 
       \and I. Mart\'i-Vidal\inst{1,2} 
       \and A. Alberdi\inst{3}
       \and M.A. P\'erez-Torres\inst{3} 
       \and E. Ros\inst{2,1}
       \and P.J. Diamond\inst{4} 
       \and J.C. Guirado\inst{1}
       \and L. Lara$^{\dag}$\inst{5}
       \and I.I. Shapiro\inst{6}
       \and C.J. Stockdale\inst{7} 
       \and K.W. Weiler\inst{8}
       \and F. Mantovani\inst{9} 
       \and R.A. Preston\inst{10}
       \and R.T. Schilizzi\inst{11} 
       \and R.A. Sramek\inst{12}
       \and C. Trigilio\inst{13} 
       \and S.D. Van Dyk\inst{14}
       \and A.R. Whitney\inst{15}}

\institute{Departamento de Astronom\'ia, Universidad de Valencia, 
           Valencia, Spain\\
           \email{J.M.Marcaide@uv.es}
        \and
           Max-Planck-Institut f\"ur Radioastronomie, Bonn, Germany
        \and
           Instituto de Astrof\'{\i}sica de Andaluc\'{\i}a, 
           CSIC, Granada, Spain 
        \and
           Jodrell Bank Observatory, University of Manchester, 
           Manchester, England
        \and
           $^{\dag}$Deceased; Universidad de Granada, Granada, Spain
        \and
           Harvard-Smithsonian Center for Astrophysics, Cambridge, MA, USA
        \and
           Marquette University, Milwaukee, WI, USA
        \and
           Naval Research Laboratory, Washington D.C., USA
        \and
           Istituto di Radioastronomia, INAF, Bologna, Italy
        \and
           Jet Propulsion Laboratory, NASA, Pasadena, CA, USA
        \and
           International SKA Project Office, Dwingeloo, The Netherlands
        \and
           National Radio Astronomy Observatory, Socorro, NM, USA
        \and
           Istituto di Radioastronomia, INAF, Noto, Italy
        \and
           Spitzer Science Center, Caltech, Pasadena, CA, USA
        \and
           Haystack Observatory, MIT, Westford, MA, USA}

\date{Sent to A\&A on 23 March 2009; accepted on 27 April 2009.}

\abstract{

We studied the growth of the shell-like radio structure of
supernova SN\,1993J in M\,81 from September 1993 to October 2003
with very-long-baseline interferometry (VLBI) observations at the
wavelengths of 3.6, 6, and 18~cm. We
developed a method to accurately determine the outer radius ($R$)
of any circularly symmetric compact radio structure such as SN\,1993J.

The source structure of SN\,1993J remains circularly symmetric
(with deviations from circularity under 2\%) over almost 4000
days. We characterize the decelerated expansion of SN\,1993J
until approximately day 1500 after explosion with an expansion
parameter $m= 0.845\pm0.005~(R \propto t^{m})$. However, from that
day onwards the expansion differs when observed at 6 and
18~cm. Indeed, at 18~cm, the expansion can be well characterized
by the same $m$ as before day 1500, while at 6~cm the expansion
appears more decelerated, and is characterized by another
expansion parameter, $m_{6}= 0.788\pm0.015$. Therefore, since
about day 1500 onwards, the radio source size has been progressively
smaller at 6~cm than at 18~cm. These findings differ significantly 
from those of other authors in the details of
the expansion. In our interpretation, the supernova expands with a
single expansion parameter, $m= 0.845\pm0.005$, and the 6~cm
results beyond day 1500 are caused by physical effects, perhaps also
coupled to instrumental limitations. Two physical effects may be
involved: (a) a changing opacity of the ejecta to the 6~cm
radiation, and (b) a radial decrease of the magnetic field in the
emitting region.

We also found that at 6~cm about 80\% of the 
radio emission from the backside of the shell behind the ejecta is absorbed
(our average estimate, since we cannot determine any possible
evolution of the opacity), and the width of the radio shell is
$(31\pm2)~\%$ of the outer radius. The shell width at
18~cm depends on the degree of assumed absorption. For $80~\%$
absorption, the width is $(33.5\pm1.7)~\%$, and for $100~\%$
absorption, it is $(37.8\pm1.3)~\%$.

A comparison of our VLBI results with optical spectral line
velocities shows that the deceleration is more pronounced in the
radio than in the optical. This difference might be due to a
progressive penetration of ejecta instabilities into the shocked
circumstellar medium, as also suggested by other authors.

}

\keywords{galaxies: individual: M81 -- radio continuum: stars 
-- supernovae: general -- supernovae: individual: SN1993J 
-- techniques: interferometric}

  \maketitle

\section{Introduction}

Supernova \object{SN\,1993J} was visually discovered in the nearby galaxy 
\object{M\,81}
on 28 March 1993 by F. Garc\'{\i}a (Ripero \& Garc\'{\i}a \cite{Ripero1993}).
It reached $m_{v}$=10.8 and became the brightest supernova in the
northern hemisphere since \object{SN\,1954A} (see Matheson et al. 
\cite{Matheson2000b} and
references therein). The relatively small distance to M\,81 (3.6
Mpc, Freedman et al. \cite{Freedman1994}) and the high northern declination of
M\,81 soon made SN\,1993J one of the best observed supernova ever,
and particularly so at very high angular resolution. Although
initially classified as Type II (Filippenko et al. \cite{Filippenko1993}),
it did not behave like other Type II -plateau or -linear
supernovae. Its light curve showed two peaks separated by about 2
weeks.

The unusual initial behavior of the light curve led many modelers
to conclude that SN\,1993J was the result of a core-collapse
explosion of a progenitor that had lost a significant fraction of
its hydrogen envelope, leaving less than one solar mass of
hydrogen in the core. Mass-loss from a massive star through winds
was proposed by H\"oflich et al. (\cite{Hoflich1993}), an
explosion of an asymptotic giant branch star of smaller main
sequence mass with a helium-rich envelope was proposed by
Hashimoto et al. (\cite{Hashimoto1993}), and stripping of hydrogen by
a companion in a binary system was proposed by many other
modelers. Models of the light curve and spectra (Nomoto et al.
\cite{Nomoto1993}; Filippenko et al. \cite{Filippenko1993}; Schmidt et al. \cite{Schmidt1993};
Swartz et al. \cite{Swartz1993}; Wheeler et al. \cite{Wheeler1993}; Podsiadlowski et al.
\cite{Podsiadlowski1993}; Ray et al. \cite{Ray1993}; Taniguchi et al. \cite{Taniguchi1993};
Shigeyama et al. \cite{Shigeyama1994}; Utrobin \cite{Utrobin1994}; Bartunov et al. \cite{Bartunov1994};
and Woosley et al. \cite{Woosley1994}) suggested ejecta masses in the range
2-6 M$_{\odot}$, with only 0.1-0.9 M$_{\odot}$ in a thin outer
hydrogen envelope with an initial radius of several hundred solar
radii. The first maximum in the optical light curve was
interpreted as being caused by shock heating of the thin envelope and the
second maximum by the radioactive decay of $^{56}$Co.

Later studies continued to suggest that a low-mass envelope of
hydrogen on a helium core was the most likely scenario for the
progenitor (Young et al. \cite{Young1995}; Patat et al. \cite{Patat1995}; Utrobin \cite{Utrobin1996};
Houck \& Fransson \cite{Houck1996}). The low-mass outer layer of hydrogen
would give the initial appearance of a Type II, but the spectrum
would slowly change to one more similar to that of a Type Ib, as
had already been considered by Woosley et al. (\cite{Woosley1987}) and
Filippenko (\cite{Filippenko1988}) for \object{SN\,1987K}. Following 
Woosley et al. (\cite{Woosley1987}), we conclude that SN\,1993J is of Type IIb.

The binary system scenario also received support from
pre-supernova photometry of the region, which indicated the
presence of more than one star (Aldering et al. \cite{Aldering1994}). According
to Filippenko et al. (\cite{Filippenko1993}), the progenitor was probably a giant
of type K0~I in a binary system. When, much later, the companion
to the progenitor was discovered (Maund et al. \cite{Maund2004}), the binary
system scenario received final backing.

Trammell et al. (\cite{Trammell1993}) and Tran et al. (\cite{Tran1997}) found
optical continuum polarization from SN\,1993J at the level of 1\%
and argued that the polarization implied an overall asymmetry,
although the source of the asymmetry was not identified. The presence
of SN\,1993J in a binary system provided a plausible source of the
asymmetry.

Models of early spectra reproduced their overall shape (Baron et
al. \cite{Baron1993}), but had difficulties fitting line strengths (Baron et
al. \cite{Baron1994}; Jeffery et al. \cite{Jeffery1994}; and Clocchiatti et al. 
\cite{Clocchiatti1995}). Wang
\& Hu (\cite{Wang1994}), Spyromilio (\cite{Spyromilio1994}), and Matheson et al. 
(\cite{Matheson2000a}) argued in favor of clumpy ejecta.

An early UV spectrum taken with the HST by Jeffery et al. (\cite{Jeffery1994})
showed a smooth spectrum similar to \object{SN\,1979C} and 
\object{SN\,1980K}, both of
which were also radio sources. Branch et al. (\cite{Branch2000}) suggested that
the illumination from circumstellar interaction might reduce the
relative strengths of line features and produce featureless UV
spectra. Indeed, the presence of circumstellar interaction could
be clearly seen in late nebular-phase spectra (Filippenko et al.
\cite{Filippenko1994}; Li et al. \cite{Li1994}; Barbon et al. \cite{Barbon1995}; Finn
et al. \cite{Finn1995}) with H$_{\alpha}$ lines beginning to dominate the
spectrum. Both Patat et al. (\cite{Patat1995}) and Houck \&
Fransson (\cite{Houck1996}) concluded that the late-time optical spectra could
only be powered by a circumstellar interaction, since radioactive
decay seemed to be insufficient.

Further support for circumstellar interaction came from the early
detection of X-rays (Zimmerman et al. \cite{Zimmerman1994}; Kohmura et al. \cite{Kohmura1994}).
Those X-rays could come from either the shocked wind material or
from the reverse-shocked supernova ejecta (Suzuki \& Nomoto \cite{Suzuki1995};
Fransson et al. \cite{Fransson1996}), according to the standard
circumstellar interaction model (SCIM) for supernovae.

The SCIM considers supernova ejecta with steep density profiles
($\rho_{ej} \propto r^{-n}$) shocked by a reverse shock that moves
inward (in a Lagrangian sense) from the contact surface, and a
circumstellar medium (CSM) with density profile $\rho_{CSM}
\propto r^{-s}$ shocked by a forward shock that moves outward from
the contact surface ($s=2$ corresponds to a steady wind). For $n >
5$, self-similar solutions are possible (Chevalier \cite{Chevalier1982b}); the
radii of the discontinuity surface, forward shock, and reverse
shock are then related, and all evolve in time with the power law
$R \propto t^{m}$, where $t$ is the time after explosion and $m$
is the deceleration parameter, which is determined by $n$ and $s$
in terms of the expression $m=(n-3)/(n-s)$. In this model, radio
emission would arise from the shocked region between the supernova
ejecta and the CSM resulting from the wind of the supernova's
progenitor star (Chevalier \cite{Chevalier1982a}).

Radio emission at 2~cm from SN\,1993J was detected within two
weeks after the explosion by Pooley \& Green (\cite{Pooley1993}) and soon light
curves were available at 1.3, 2, 3.6, 6, and 20~cm (Van Dyk et al.
\cite{VanDyk1994}). The high level of radio emission from this supernova and
its high northern declination paved the way for a superb sequence
of very-long-baseline interferometry (VLBI) observations, which
started very early on (Marcaide et al. \cite{Marcaide1994}, Bartel et al. \cite{Bartel1994})
and have continued for over a decade. From VLBI observations,
Marcaide et al. (\cite{Marcaide1995b}) found a spherically symmetric shell of
width about 0.3 times the outer radius, and Marcaide et al.
(\cite{Marcaide1995a}) showed the first movie of the self-similar growth of the
shell over one year. With VLBI data from three years of
observations, Marcaide et al. (\cite{Marcaide1997}) reported deceleration in the
expansion of the shell and estimated a value of the deceleration
parameter $m=0.86 \pm 0.02$. Combining this estimate with the
determination of the opacity due to free-free absorption in the
CSM by Van Dyk et al. (\cite{VanDyk1994}), we derived a value of
$s=1.66^{+0.12}_{-0.25}$ (Marcaide et al. \cite{Marcaide1997}) in agreement with
the value $s=1.7$ given by Fransson et al. (\cite{Fransson1996}) to explain the
X-ray emission. However, such a determination of the free-free
opacity, and hence of the value of $s$, has been questioned by
Fransson \& Bj\"ornsson (\cite{Fransson1998}) who instead argued in favor of
$s=2$ and emphasized the importance of synchrotron
self-absorption. P\'erez-Torres et al. (\cite{PerezTorres2001}) also emphasized the
importance of synchrotron self-absorption in the interpretation of
the radio light curves.

The determination of the deceleration parameter also allows for a
direct comparison of ejecta density profiles determined from
modelling the emission spectrum. Using NLTE
algorithms, Baron et al. (\cite{Baron1995}) derived a value of $n=50$ shortly 
after the explosion, decreasing to $n=10$ at late epochs. The later $n$ corresponds
(for $s=2$) to $m=0.875$, compatible with the determination by
Marcaide et al. (\cite{Marcaide1997}). Additionally, such a value of $m$ is
compatible (for the assumed distance of 3.6 Mpc to M\,81) with the
expansion speeds of 14,000 km $s^{-1}$ up to ~1000 days after
explosion (Garnavich \& Ann \cite{Garnavich1994}) and of 10,000 km $s^{-1}$ for
days 1000-1400 after explosion (Fransson et al. \cite{Fransson2005}). The radio
spectrum at long wavelengths has been studied by P\'erez-Torres et
al. (\cite{PerezTorres2002}), and Chandra et al. (\cite{Chandra2004}). Fransson \& Bj\"ornsson
(\cite{Fransson1998}) proposed a model in which the size of the radio emitting
region would be discernibly wavelength dependent. Those
long-wavelength results and the Fransson \& Bj\"ornsson model will
be considered in Sect. \ref{synchmeanlife}.

The expansion of the radio shell has taken place with remarkable
spherical symmetry (Marcaide et al. \cite{Marcaide1995a}, \cite{Marcaide1997}, this paper;
Bietenholtz et al. \cite{Bietenholz2001}, \cite{Bietenholz2003}, \cite{Bietenholz2005}; 
Alberdi \& Marcaide \cite{Alberdi2005}). This
result, while complying nicely with the simplest SCIM, is in sharp
contrast with the claims of asymmetry (Trammell et al. \cite{Trammell1993}; Tran
et al. \cite{Tran1997}) based on the detection of optical polarization from
the supernova ejecta, which require a ratio of about 0.6 for the
radii of an elliptical emission model. It is hard to imagine that
the ejecta could have such an asymmetry and the outer shock front,
as delineated by the outer surface of the radio emission, such a
remarkable symmetry. These characteristics would appear to be
inconsistent with the SCIM. Perhaps, as pointed out by Matheson et
al. (\cite{Matheson2000b}), there is no such inconsistency between the early
optical polarimetric observations and the VLBI observations since
the two types of observations may probe different regions of the
supernova shell.

Great effort has been invested in determining the width of the
expanding radio shell and the value of $m$ as a function of time
after explosion by two groups working on independently acquired
VLBI data. Each group has used different data acquisition and
analysis strategies. Bartel et al. (\cite{Bartel2002}) confirmed the
deceleration reported earlier by Marcaide et al. (\cite{Marcaide1997}), but
claimed that the values of $m$ differ for different
expansion periods. Those results were in agreement with previous
results from numerical simulations made by Mioduszewski et al.
(\cite{Mioduszewski2001}) using a rather specific explosion model. Preliminary
observational evidence to the contrary was provided by Marcaide et
al. (\cite{Marcaide2005a}) and definitive evidence is provided in this paper.
After the initial estimate by Marcaide et al. (\cite{Marcaide1995b}) of a shell
width of $0.3\pm0.1$ times the size of the outer radius of the
source, Bartel et al. (\cite{Bartel2000}) reported shell widths as narrow as
$0.205 \pm0.015$. However, after Bietenholz et al. (\cite{Bietenholz2003}), and
Marcaide et al. (\cite{Marcaide2005b}) provided evidence of absorption in the
central part of the shell emission, Bietenholz et al. (\cite{Bietenholz2005})
revised their estimates of the shell width to $0.25 \pm 0.03$,
consistent with the value reported by Marcaide et al. (\cite{Marcaide1995b}), and
closer to, but still inconsistent with, the more accurate estimate
reported in this paper.

The study of SN\,1993J has been very important for at least two
reasons: a) for the first time a clear transition from Type II to
Type Ib was observed, thus linking Type Ib (and for that matter
Type Ic) to massive core collapse supernovae, such as Type II, rather
than the thermonuclear explosion supernovae, such as Type Ia; and b)
for the first time a long sequence of images following a supernova
were obtained to provide detailed information on the expansion
rate. The results from such monitoring have lent support to the
SCIM initially proposed by Chevalier (\cite{Chevalier1982a}, \cite{Chevalier1982b}). 
However, Bartel et al. (\cite{Bartel2002}) claim to have detected departures 
from a self-similar
expansion with regimes of changing expansion rates over different
periods. In this paper, we provide evidence contrary to such
claims based on our own data and on the use of new analysis tools,
and support the validity of the SCIM model in which additional
fine observational effects have to be taken into account.

The remainder of this paper is organized as follows: We first
describe our observations. Then we describe a novel approach to
improved imaging and measuring of source size and shell width, and
compare the results obtained with different methods. We give
tentative physical and observational reasons for the rather
surprising result that the apparent expansion rate at two
wavelengths is slightly, but significantly, different. Finally,
assuming the distance to SN\,1993J as that obtained by Freedman et
al. (\cite{Freedman1994}) for M\,81, we compare radio and optical results.

\section{Observations, correlation, and data reduction}

Table \ref{table-all} summarizes all our VLBI observations of
SN\,1993J at 3.6, 6, and 18\,cm from 1993 September 26 through 2003
October 17. Our early results at 3.6 and 6\,cm were published by Marcaide
et al. (\cite{Marcaide1995a}, \cite{Marcaide1995b}, \cite{Marcaide1997}). 
In this paper, we reanalyze those
published data together with the new data using new analysis
methods that will be described below.

The antennas that participated in all or some of our observations
are: the VLBA (10 identical antennas of 25\,m diameter each spread
over the US from the Virgin Islands to Hawaii), the phased-VLA
(equivalent area to a paraboloid of 130\,m diameter, New Mexico,
USA), the Green Bank Telescope (100\,m, WV, USA), Goldstone (70\,m,
CA, USA), Robledo (70\,m, Spain), and the European VLBI Network
including Effelsberg (100\,m, Germany), Medicina (32\,m, Italy),
Noto (32\,m, Italy), Jodrell Bank (76\,m, UK), Onsala (20 and 25\,m,
Onsala, Sweden), Westerbork (equivalent area to a paraboloid of 93\,m 
diameter, The Netherlands). The Goldstone and Robledo antennas
could only take part in the  3.6 and 18\,cm observations.
Westerbork only took part in the 6 and 18\,cm observations. The
effective array consisted typically of about 15 antennas. The
recording was each time set to the maximum available rate at that
time (256\,Mbps), 2-bit sampling, single polarization mode (RCP at
3.6\,cm and LCP at 6 and 18\,cm). The synthesized bandwidth was 64\,MHz 
(except at the VLA, where it was limited to 50\,MHz). The data
were correlated either at the Max Planck Institut f\"ur
Radioastronomie, Bonn, Germany, when the MkIV recording system was
used (see Marcaide et al. \cite{Marcaide1997}), or at the National Radio
Astronomy Observatory, Socorro, NM, USA, when the VLBA recording
was used [after 1997 February 2]. We provide in Table
\ref{table-all} a summary of the VLBI observations including the
rms noise of each of the reconstructed maps of SN\,1993J.

A typical 12h observation cycled between SN\,1993J and the core of
M\,81, and observed occasionally \object{0917+624} and 
\object{0954+658}.
Additionally, we observed a number of sources such as \object{3C\,286} 
and \object{3C\,48} at appropriate times during each observing session 
for flux-density calibration purposes. Once the correlated data were
available, we initially calibrated the data with the radiometry
information obtained at each antenna participating in the array.
For all data reduction purposes apart from mapping, we used the
NRAO AIPS package, and for mapping we used DIFMAP (Shepherd et al.
\cite{Shepherd1995}).

We usually started the data reduction by analyzing the 0917+624
and 0954+658 data. We used these two sources for instrumental
calibration (we ran program FRING on data integrated over the
duration of a scan to determine the residual delays which aligned
the 16 channels of the IF for 0917+624 and 0954+658.)
After previously reducing the residual fringe 
rates to a weighted mean of
zero, we applied those residual delays to data from the whole
observing session and, in particular, searched for new residual
phase-delay and residual delay-rate solutions for the core of
M\,81, integrating the data over the duration of each scan and
assuming a centered point model for M\,81. Finally, we mapped the
core of M\,81 in DIFMAP using, whenever necessary, phase and
amplitude self-calibration. With the map of the core of M\,81 at
hand, we used it as input to programs FRING and CALIB of AIPS to
improve the fringe search solution by removing the
contribution of the phases due to the structure of the core of
M\,81 from the data stream, and to improve the amplitude
calibration, respectively. The new fringe solution for the
complete data set (and for SN\,1993J, in particular) is thus
referred to the reference point chosen in the core of M\,81. We
finally time averaged the SN\,1993J data over 2 minutes and frequency
averaged over the synthesized band.

\begin{small}

\begin{table*}[!th]
\begin{minipage}[t]{17cm}
\caption{Summary of VLBI observations.}
\label{table-all}
\centering
\renewcommand{\footnoterule}{}
\begin{tabular}{crcccc|l|l|l}
\hline\hline

Date   & Age\footnote{Age is in days after explosion} & Map Flux\footnote{Map Flux 
corresponds to total map flux density} & Map Peak\footnote{Maximum flux density per 
unit beam in the maps} & Noise rms\footnote{Root-mean-square of the corresponding residual
maps} & Radius\footnote{SN\,1993J radius (and its standard deviation) as estimated using 
the Common-Point Method described in the text (see Sect. \ref{CPM-Sec} and Appendix 
\ref{CPM-App} for details)} 
&\multicolumn{3}{c}{$\lambda$\footnote{Observing wavelength} (cm)}\\

(dd/mm/yy) & (days) & (mJy) & (mJy beam$^{-1}$) & (mJy beam$^{-1}$) & (mas) & 3.6 & 6  & 18 \\

\hline

26/09/93 &  182 & 78.50 & 8.91 & 0.037 & 0.488 $\pm$ 0.004 &X& &  \\
22/11/93 &  239 & 57.30 & 7.32 & 0.089 & 0.628 $\pm$ 0.014 &X& &  \\
20/02/94 &  329 & 51.00 & 6.52 & 0.037 & 0.818 $\pm$ 0.015 &X& &  \\
29/05/94 &  427 & 41.50 & 5.24 & 0.094 &  1.02 $\pm$ 0.02  &X& &  \\
20/09/94 &  541 & 53.40 & 7.01 & 0.097 &  1.15 $\pm$ 0.03  & &X&  \\
23/02/95 &  697 & 44.30 & 5.53 & 0.090 &  1.48 $\pm$ 0.04  & &X&  \\
11/05/95 &  774 & 41.80 & 5.63 & 0.083 &  1.66 $\pm$ 0.04  & &X&  \\
01/10/95 &  917 & 32.20 & 4.36 & 0.170 &  1.92 $\pm$ 0.04  & &X&  \\
28/03/96 & 1096 & 31.30 & 3.47 & 0.082 &  2.21 $\pm$ 0.03  & &X&  \\
17/06/96 & 1177 & 26.50 & 2.94 & 0.110 &  2.31 $\pm$ 0.04  & &X&  \\
22/10/96 & 1304 & 26.10 & 3.12 & 0.074 &  2.61 $\pm$ 0.02  & &X&  \\
25/02/97 & 1430 & 24.44 & 2.46 & 0.085 &  2.81 $\pm$ 0.05  & &X&  \\
21/09/97 & 1638 & 21.75 & 2.52 & 0.061 &  3.09 $\pm$ 0.03  & &X&  \\
18/02/98 & 1788 & 21.28 & 2.28 & 0.061 &  3.37 $\pm$ 0.03  & &X&  \\
30/05/98 & 1889 & 22.24 & 2.94 & 0.059 &  3.48 $\pm$ 0.04  & &X&  \\
23/11/98 & 2066 & 20.62 & 2.30 & 0.045 &  3.74 $\pm$ 0.04  & &X&  \\
30/11/98 & 2073 & 39.84 & 4.67 & 0.050 &  3.84 $\pm$ 0.06  & & & X\\
10/06/99 & 2265 & 18.05 & 1.89 & 0.053 &  4.06 $\pm$ 0.04  & &X&  \\
22/09/99 & 2369 & 17.02 & 2.03 & 0.043 &  4.20 $\pm$ 0.04  & &X&  \\
06/06/00 & 2627 & 15.87 & 2.19 & 0.066 &  4.48 $\pm$ 0.06  & &X&  \\
20/11/00 & 2794 & 27.72 & 3.34 & 0.040 &  4.98 $\pm$ 0.07  & & & X\\
24/11/00 & 2798 & 14.24 & 1.64 & 0.047 &  4.71 $\pm$ 0.10  & &X&  \\
14/02/01 & 2880 & 13.10 & 1.37 & 0.057 &  4.84 $\pm$ 0.07  & &X&  \\
18/11/01 & 3157 & 10.58 & 1.28 & 0.048 &  5.21 $\pm$ 0.10  & &X&  \\
07/11/02 & 3511 & 10.36 & 1.11 & 0.047 &  5.67 $\pm$ 0.10  & &X&  \\
17/11/02 & 3521 & 16.12 & 1.80 & 0.050 &  6.06 $\pm$ 0.05  & & & X\\
29/10/03 & 3867 & 09.47 & 1.25 & 0.046 &  6.15 $\pm$ 0.11  & &X&  \\

\end{tabular}
\end{minipage}
\end{table*}

\end{small}

\section{Imaging of SN\,1993J}
\label{Imaging-Sec}

Once the SN\,1993J data had been calibrated in AIPS as described
in the previous section, the mapping of SN\,1993J was completed in
DIFMAP. Especial care was taken during the imaging process to
avoid introducing any bias that might affect the final estimate of
the SN\,1993J radius. Data that had been analyzed earlier, and the
results  already published (Marcaide et al. \cite{Marcaide1995a}, \cite{Marcaide1995b}, 
\cite{Marcaide1997}), were also reanalyzed with this new approach.

Since the structure of SN\,1993J is very circularly symmetric, the
visibilities, when referred to the center of the map, are such
that the imaginary part cancels out. All the information is then
contained in the real part of the visibility. We used this
property to determine the position of the center of the map
with data from day 1889 after explosion (i.e., 1998 May 30; see
Table \ref{table-all}). We chose this map because of its very high
quality. Its center, determined with respect to the core of M\,81
for this epoch, was used for every epoch. For each epoch, we found
that the imaginary parts of the visibilities were zero. Since this
condition was satisfied for every epoch, we concluded that
the structure remained circularly symmetric and that the center of
the structure remained stationary with respect to the core of
M\,81.

\begin{figure}[!t]
\centering
\includegraphics[height=4cm]{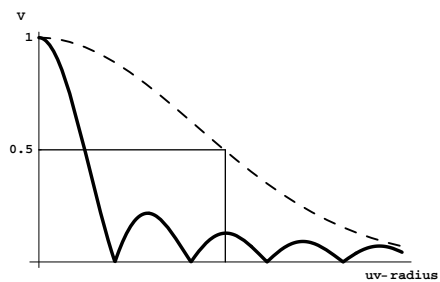}
\caption{Schematic representation of the taper used in our mapping
and model-fitting procedures. The dark continuous line represents
the amplitude of the visibilities, while the dashed one represents
the taper. The point indicated on the dashed line corresponds to
the value of the uv-radius such that the taper value is equal to
0.5. We have chosen this value to correspond to the middle of the
third lobe of the visibility amplitudes. As the supernova expands,
the lobes will shrink in the uv-plane as will the taper function,
thus increasing the width of the dirty beam in the same proportion
as the radius.} \label{fig31}
\end{figure}

To be consistent with the use of a dynamic beam, which was
introduced by Marcaide et al. (\cite{Marcaide1997}) (see also next 
section), to
avoid a bias in the measurement of the supernova expansion, a
similar use of a dirty dynamic beam had to be made during the
imaging process. This use was achieved by tapering the data with a
Gaussian taper, whose width evolved inversely with the source size.
In particular, taking advantage of the azimuthal symmetry of the
source, we used at all epochs a normalized Gaussian taper such
that its half value falls at the middle of the third lobe of the
visibility function, as shown schematically in Fig. \ref{fig31}.

Another innovation of our imaging process is the use of a
non-point source initial model in the mapping process.
Customarily, a point source model is taken as the initial model
for imaging of radio sources. However, such a choice is not the
best one for the mapping of SN\,1993J. Instead, we have taken
advantage of SN\,1993J being rather 
circularly symmetric to design a procedure that is objective and 
very useful in the fine calibration of the visibilities in the imaging
process. We now describe this procedure in detail.

Due to the circular symmetry of the source and its relatively
sharp edge, the visibilities display very clear lobes and the
details of the source structure are most evident in the second and
higher order lobes. Indeed, the Fourier transform of a perfectly
circularly symmetric source will be real. In such a case, the
phases will alternate between 0 and $\pi$ radians in the lobes,
being zero for the first lobe. We can use this circumstance to our
benefit in the case of SN\,1993J since, as noted above, it is
rather circularly symmetric, with irregularities being of a small
scale perceptible only in the higher-order lobes. Thus, if we can
guess the location of the transition between the first and second
lobes, then, using only the data from those two lobes, we can
adopt for self-calibration a perfectly symmetric model, which has
its first-to-second-lobe transition at roughly the same point as
for the data. We can thus use the data that correspond to the
first two lobes for the initial phase self-calibration, ignoring
the data corresponding to higher resolution. This self-calibration
with the program SELFCAL will force the data to have 0 phase for
the first lobe and $\pi$ for the second. Given that the solutions
obtained will be antenna dependent, a new self-calibration step,
now using all the data, will clearly define the locations of the
remaining phase changes (from 0 to $\pi$, or viceversa) for higher
resolution data.

We note that the quality of the initial guess in the location of
the transition between the first and second lobes, and the use of
different models for the initial self-calibration, are not crucial
to the procedure. The former is true because near the null the
phases are ill defined in all cases, while the latter is true
because the procedure has to do with phases, and is therefore
insensitive to the amplitudes of the model. We verified the
correctness of the previous assertions. In practice, we used a
simple (uniformly bright) disc model for the initial
self-calibration of the data in the first two lobes.

After self-calibrating the phase data, we proceeded in the usual
way of mapping, using iterative low gain CLEANing and phase
self-calibration a number of times. As a final step, we
applied
amplitude and phase calibration for every 30 minutes of data.

\begin{figure*}[!t]
\centering
\includegraphics[width=15cm,angle=0]{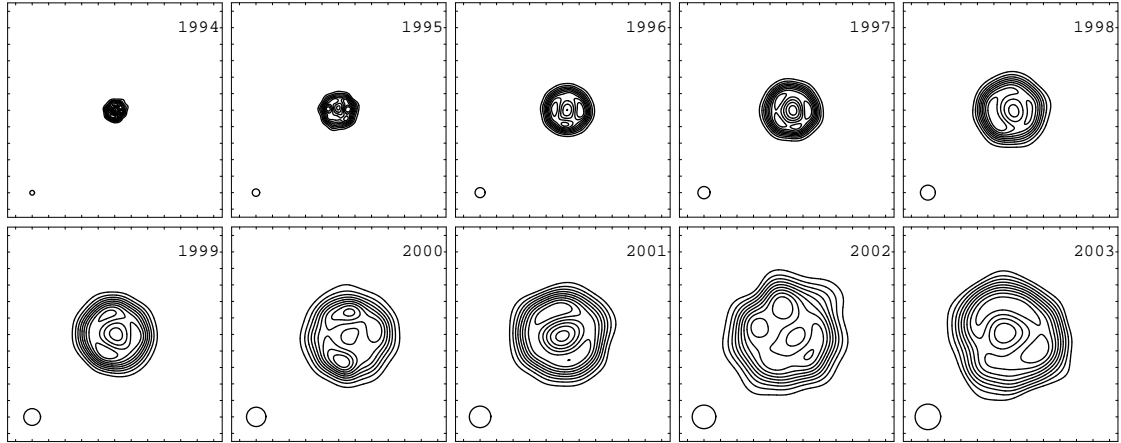}
\caption{Maps of SN\,1993J at 6\,cm corresponding to epochs in or
near October every year from 1994 through 2003. The FWHM of the
circular beam used to reconstruct each map is shown in the lower
left corner. Contours correspond to (10,20,30,40,50,60,70,80,90)\%
of peak emission (see Table 1). Tick marks are in milliarcseconds
(mas). One mas in each map corresponds to approximately 3500~AU.}
\label{fig32}
\end{figure*}

In Fig. \ref{fig32}, we show the contour maps for observations
made in or near October of every year. These maps are
representative of all the maps we have reconstructed\footnote{Color 
maps can be found in the following 
web page: \texttt{http://www.uv.es/radioastronomia/SN1993J-10yr-AA09.jpg}}.

\section{Measurement of the radius of SN\,1993J}

\subsection{Introduction}

The circular shape of the images of SN\,1993J facilitates the task
of defining a radius. Even so, since an accurate value of the
radius of SN\,1993J at each epoch is crucial to study the details
of the expansion, especial care has to be taken in the estimate of
the radius. Before we describe our present approach, we note that
Marcaide et al. (\cite{Marcaide1997}) and Bartel et al. (\cite{Bartel2000}) took different
approaches to estimating the radius in their attempts to determine
the details of the expansion. Marcaide et al. (\cite{Marcaide1997}) used the
average of the radial distances from the map center to the 50 \%
contour level of the maps for a number of directions. To
avoid a bias in the radii estimates, these authors had convolved
the source models with beams proportional to supernova sizes to
obtain the maps. Instead, Bartel et al. (\cite{Bartel2000}) estimated the
supernova outer radius (as well as the inner radius) by fitting a
shell model in Fourier space. For this purpose, they assumed a
specific, spherical, optically-thin source model.

In this work, we tried to overcome the drawbacks of each of
the previous measuring schemes. The principal drawback of the
procedure of Bartel et al. (\cite{Bartel2000}) was that the estimates were
model-dependent. This drawback became apparent when it was found
later that the emission from the central part of the source is
greatly suppressed (Bartel et al. \cite{Bartel2002}, Marcaide et al. \cite{Marcaide2005b}). We
refined the procedure of Marcaide et al. (\cite{Marcaide1997}) by developing
new tools that allow for accurate measurements on the sky plane,
while keeping the measuring scheme model-independent.

Before describing our method in detail, we illustrate it with two
simple one-dimensional cases: (a) We consider a uniform source whose
emission intensity is nonzero for all $r \leq R$ and zero for $r > R$
(that is, the source is the equivalent of a disk in 2 dimensions).
We convolve this step-function source with Gaussian 
beams of different widths, $\sigma$, such that in every case $\sigma$ is
much smaller than $R$. After convolution, all resulting functions
will cross at $R$ at half the height of the step function. Thus, the
crossing point of the resultant functions exactly determines the
source radius $R$. (b) Secondly, we consider a narrow ``boxcar'' 
source, that is, a uniform source whose emission is nonzero only over a narrow
region just short of its outer edge, $r = R$ (that is, the equivalent
to a thin shell in 2 dimensions). If we convolve this model of the
source with the Gaussian beams $\sigma$ described above, we
find that the center of the resultant function will almost
coincide with $R$, but the position of the outer half-power point
will be larger than $R$.

For a one-dimensional model in-between the above extreme models
(e.g., a ``boxcar'' of width comparable to the values of $\sigma$ 
(that is, the equivalent of a thick shell in 2 dimensions)), 
using the outer half-power point of the resultant function to determine the
radius R would give a result r$>$R. The ratio r/R would remain
constant provided the model maintained its functional shape while
changing R (self-similar change) and provided $\sigma$ changed
fractionally the same amount as R.

This latter, intermediate, one-dimensional model illustrates the
idea of the Marcaide et al. (\cite{Marcaide1997}) method: use of dynamical beams
to reconstruct the SN\,1993J images before measuring their sizes
at the 50\% contour level. While each of the size measurements
might be slightly biased, the expansion measured will not be
biased provided that the shape of the source emission does not change
with time. Since the central absorption (found later) in SN\,1993J
appears to have been strong at all times, the expansion results
given by Marcaide et al. (\cite{Marcaide1997}) are likely to be nearly 
unbiased. The expansion measured with Bartel et al.'s method (fitting a model to
the visibilities) is likely to be biased since use of an incorrect
model (optically thin, without central absorption) will bias each
measurement of the radius, and likely in a time-dependent manner
as the amount of the visibility sidelobes involved in the fit
changes as the source grows in size. The Common Point Method
described in the next section has, in this respect, the same
advantages as the method used by Marcaide et al. (\cite{Marcaide1997}) but is
more accurate.

\subsection{The Common Point Method}
\label{CPM-Sec}

\begin{figure}[!t]
\centering
\includegraphics[width=8cm]{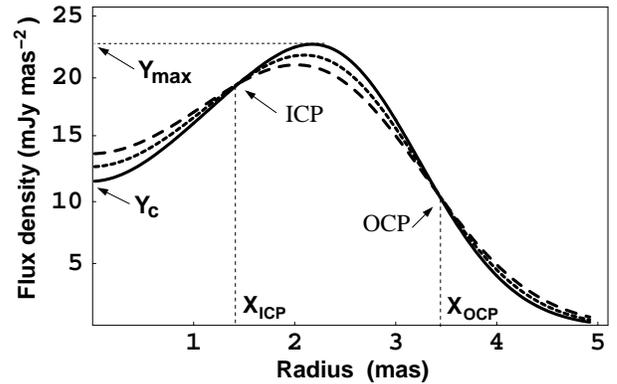}
\caption{(Solid line) Profile obtained azimuthally averaging the
map obtained at 6\,cm from observations made on day 1889 after
explosion (see Fig. \ref{fig41}(a) in Appendix \ref{CPM-App}) for a convolving
beam size of 1.74 mas. Tick marks are in mas (X axis) and in mJy
mas$^{-2}$ (Y axis). $Y_{max}$ is the value of the profile at
maximum and $Y_{c}$ is the value of the profile at the source
center. The dotted and dashed lines are profiles obtained in the
same way but using convolving beam sizes of 1.92 and 2.1 mas,
respectively. The three profiles cross at two points; the position
of the outer one ($X_{OCP}$) is taken to be the source size in the
CPM; the difference between it and the inner one ($X_{OCP}$ -
$X_{ICP}$) is taken to be related to the shell width of the
source.} \label{fig42}
\end{figure}

Given a map of SN\,1993J, if we azimuthally average its brightness
distribution we obtain a profile similar to that shown in Fig.
\ref{fig42} (solid line). For maps corresponding to the same model
but reconstructed with beams of different sizes, different
profiles will be obtained. However, if those profiles are
superimposed to each other we find that they cross 
approximately at two points, as
shown in Fig. \ref{fig42}. We call ``outer common point'' (OCP)
the outer approximate common point for all profiles. (We take the
name of the method from this characteristic.) We use the radial
distance of the OCP, $X_{OCP}$, as an estimate of the source
radius. We call ``inner common point'' (ICP) the inner approximate
common point for all profiles. The position of the ICP, $X_{ICP}$,
is related to the inner shell border.

In practice, we reconstruct a map using a beam of size (full width
at half maximum, FWHM) equal to half the source radius, $X_{OCP}$,
and iterate until the estimate of the source radius changes
fractionally in an iteration by less than 0.01; three iterations
are usually sufficient to determine the source size. The final
estimate of $X_{OCP}$ is our CPM estimate of the source size. The
uncertainty in this estimate is assumed to be related to the lack
of circularity of the source (see Appendix \ref{CPM-App}).

The whole method relies strongly on the properties of the outer
common point. Because of this reliance, we included in
Appendix \ref{CPM-App} a mathematical description of the method
and details about how the method works in practice.

\subsection{Considerations on the use of the Common Point Method}
\label{Considerations-Sec}

Simulations can be used to investigate the biases in the size
determination with the CPM. The number of simulations can be
reduced considerably by taking into account what is already known
about the source emission structure. Previous VLBI observations
provided strong support to a shell-like structure of the
emission, although there is not as yet agreement on the size of
the shell width (Marcaide et al. \cite{Marcaide1995a}; Bartel et al. \cite{Bartel2000}). 
There is also evidence that part of the emission from an otherwise
optically thin shell is suppressed (Bartel et al. \cite{Bartel2002}; Marcaide
et al. \cite{Marcaide2005b}); it appears that the emission from the part
of the shell behind the ejecta is very much absorbed. Hence, we
considered two classes of models to test the accuracy and
bias of the CPM. One class of models consists of an optically thin
shell, and the other consists of an optically thin shell with the
emission from behind the ejecta suppressed by absorption. Furthermore,
for each class we  considered in our simulations 3 different
shell widths: 0.25, 0.30, and 0.35 times the outer radius, $R$.
The relevant information in each case can be extracted by
considering 4 significant points in profiles such as those in 
Fig. \ref{fig42}: $Y_{max}$, the value of profile at maximum; $Y_{c}$,
the value of the profile at the source center; $X_{ICP}$, the
radial position of the inner common point; and $X_{OCP}$, the
radial position of the outer common point.

$X_{OCP}$ is our estimate of the model radius. Is this
estimate unbiased? No. It is biased by a factor that is model
dependent. Table \ref{table-bias} shows the ratio of $X_{OCP}$ 
and the true radius for each model in our simulations. Two more ratios
carry significant information about the model structure: the ratio
$\rho_{1} = \frac{Y_{c}}{Y_{max}}$ gives an indication of the
absorption. The larger the absorption the smaller $Y_{c}$ will be.
Given that $Y_{max}$ is less sensitive than $Y_{c}$ to the
absorption, a smaller $Y_{c}$ implies a smaller $\rho_{1}$. The
ratio $\rho_{2} = \frac{X_{ICP}}{X_{OCP}}$ gives an indication of
the shell width.

\begin{table}[!t]
\begin{minipage}[t]{\columnwidth}
\caption{Biases in the size determination.}
\label{table-bias}
\centering
\begin{tabular}{cc|c}
$\xi$\footnote{$\xi$ refers to the shell width in units of the source
radius (fractional shell width) of the model.} 
& Absorption\footnote{Absorption refers to a blockage of all the 
emission coming from the part of the shell behind the ejecta (see text).} 
 & $R/X_{OCP}$~\footnote{Ratio of model source size, $R$, to size determined 
with the CPM, $X_{OCP}$.} \\
\hline
0.35 & YES & 0.975 \\
0.30 & YES & 0.970 \\
0.25 & YES & 0.961 \\
\hline
0.35 & NO & 1.012 \\
0.30 & NO & 0.995 \\
0.25 & NO & 0.977 \\
\end{tabular}
\end{minipage}
\end{table}

\subsection{Measurements in Fourier space}
\label{Fourier-Sec}

In the next section we present our results obtained with the
CPM. To compare our results with those of other researchers, obtained 
using a Fourier analysis, we also analyzed our
data in Fourier space. For completeness, we describe the
Fourier analysis scheme that we used.

Since the imaginary part of the visibility was always nearly zero,
because of the circular symmetry of the source structure 
and to our choice of the phase center, as explained in Sect.
\ref{Imaging-Sec}, we used only the real part  in fitting. The
data were weighted using a taper (shown in Fig. \ref{fig31}) to
downweight the data from noisy long baselines and avoid any
significant change in bias as the source size increased. The model
used to fit the data was an optically thin shell with total
suppression of the emission from behind the ejecta (see Appendix
\ref{Modelfit-App}). The free parameters in the model-fitting were
the source's total flux density, the source's radius and the
shell's width. Since in our case the real part of the visibility
also has circular symmetry, we azimuthally averaged the data in Fourier
space to increase the SNR of the data. The averaging is
made using bin sizes that scale inversely with the source size
and thus always sample the visibility in the same manner (see
Appendix \ref{Modelfit-App} for details).

A simultaneous fit to source radius and shell width does not
usually yield estimates with low uncertainties and low
correlations between fitting parameters in cases of low flux
density and relatively poor UV-coverage (i.e., $\chi^{2}$ does not
then have a sharply defined minimum). In these cases, we fixed one
parameter and estimated the other. First, we fixed the shell width
(to values that will be given below) and fit the source radius, and
later we fixed the source radius to the value obtained with the
CPM and fit the shell width. We used the Levenberg-Marquardt
(e.g., Gill \& Murray \cite{Gill1978}) non-linear least-squares technique, as
implemented in the Mathematica 5.0 Package (Wolfram \cite{Wolfram2003}).

\section{Expansion of SN\,1993J}
\label{Expan1-Sec}

\begin{figure*}[!t]
\centering
\includegraphics[height=16cm, angle=-90]{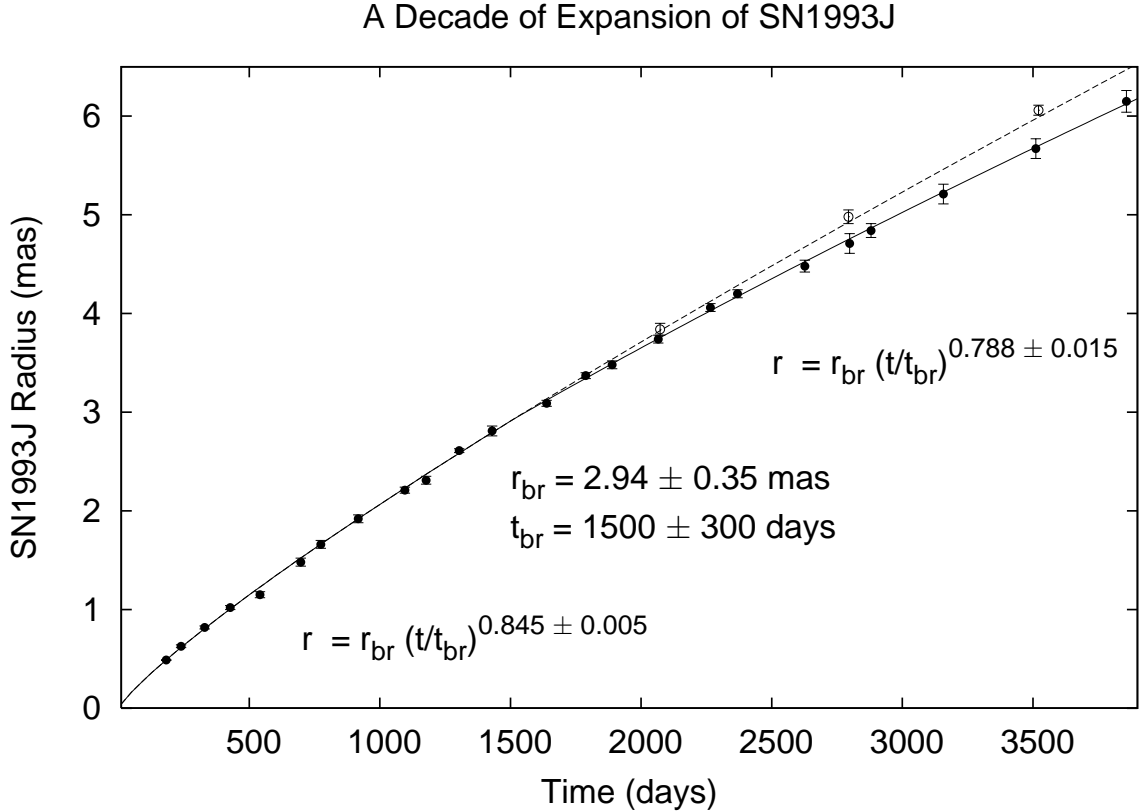}
\caption{Expansion of SN\,1993J over a decade, measured from the
estimated date of explosion. Filled circles represent 3.6 and 6~cm
data and empty circles 18\,cm data. The continuous line corresponds
to a model in which two power laws, one before and one after a
break point, $ t_{br}$, are fit to the data.  The model
predictions from the power law before $ t_{br}$ have been
extrapolated beyond the break (dashed line) to compare with the
18\,cm data which were not used in the fit.} \label{fig51}
\end{figure*}

The analysis of all the images of supernova SN\,1993J with the CPM
yields for the supernova radius the results shown in Table
\ref{table-all}. Figure \ref{fig51} plots those results against
time elapsed since the explosion. We also show in Fig. \ref{fig51} a
single fit to both the 3.6 and 6\,cm, but not the 18~cm data. The
fit shown was obtained using the fitting procedure implemented in
the program Gnuplot (Mathematica gave the same result). Our
weighted least squares fit of the supernova radius $R$ as a function
of time has 4 parameters since we used the functional form $R
\propto t^m$, and allowed for two regimes of expansion, each with
its own value for $m$. We thus estimated two values of $m$, the
epoch of transition between these two regimes (``break-time''), and
the radius of the supernova at this break-time, $ t_{br}$. The
reduced chi-square of the fit, for the standard errors estimated
with the CPM is 0.1. Hence, we divided each of these standard
errors by the square-root of ten to obtain a reduced chi-square of
unity. The estimates in Col. 6 of Table \ref{table-all} are
shown with these re-scaled uncertainties. These estimates and
standard errors, which in part account for the departures from
circularity as explained above, indicate that the radio supernova
image remains circularly symmetric over ten years with departures
from circularity at the level of 2\% or less (see Appendix
\ref{CPM-App}.3 for details.)

The data at 3.6\,cm were only available at early epochs (see Table
\ref{table-all}). Thus, it is only in the 6\,cm data that we see a
break in the expansion rate. The 3.6\,cm data in Table
\ref{table-all} are consistent with the 6\,cm data. In Fig.
\ref{fig51}, the dashed line indicates an extrapolation with the
time dependence determined before the break. It is remarkable that
the 18\,cm data are consistent with such an expansion in sharp
contrast to the data at 6\,cm, which require a significantly
different expansion rate. The ratio of the sizes at 18\, cm
to those at 6\,cm
thus evolves after the break as $\propto t^{0.057 \pm 0.016}$ . By
day 3500 after the explosion, the discrepancy between the size estimate
at 6\,cm and 18\,cm is about 0.4\,mas, i.e., about 7\%.

We also modelled the expansion curve obtained by 
applying the CPM to the phase-referenced images of the supernova, (i.e., 
without any self-calibration) and the results obtained are totally 
compatible with those using self-calibrated images. We obtain 
$m_1 = 0.845 \pm 0.007$, $m_2 = 0.799 \pm 0.020$, and 
$t_{break} = 1500 \pm 400$. The scatter in the data 
around the expansion curve and the parameter 
uncertainties are larger. 

The use of size estimates from model fitting to the 
visibilities (using the  model described in Appendix B with the fractional 
shell width fixed to 
0.3) also results in fitted parameters which are very similar. In 
this case, we obtain $m_1 = 0.88 \pm 0.05$, $m_2 = 0.799 \pm 0.017$, and 
$t_{break} = 2250 \pm 300$. We notice again that the parameter 
uncertainties are in this case also larger than those obtained using 
the CPM with the self-calibrated images.

\begin{figure}[!t]
\centering
\includegraphics[height=5cm]{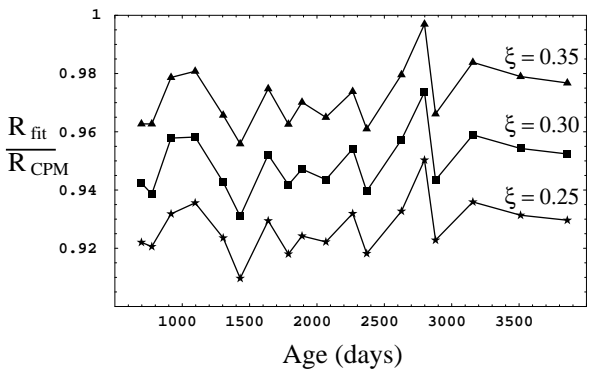}
\caption{Comparison of estimates of the supernova radius at 6~cm
obtained using the CPM and a fit to the visibilities with the
model explained in the text. The ratio of these estimates is shown
for 3 different fractional widths, $\xi$, of the shell model.}
\label{fig52}
\end{figure}

We repeated this procedure using the values 0.25 and 0.35 for the 
fractional shell width. The results for the three fractional shell 
widths are shown in Fig. \ref{fig52}, normalized to the estimates 
obtained with the CPM. The results shown in the figure are further 
evidence of the consistency of the results obtained with all methods.

For each value of the shell width of the source model, the ratio
$R_{fit}/R_{CPM}$ remains rather constant, showing a scatter with
a fractional standard deviation of only about 1\%. The
significance of the different values of the ratio will be
discussed in Sect. \ref{Central-Abs-Sec}. Here, we note only that this 
constancy in the ratio is tantamount to a replication via fitting of the
expansion characteristics determined with the CPM and shown in
Fig. \ref{fig51}. We should add that the CPM determines a
smoother expansion than the method based on model fitting, since
the scatter in the expansion determined with CPM estimates is
smaller than with model fitting estimates (the unweighted reduced
$\chi^{2}$ of the fit of the supernova expansion using CPM
estimates is 17\% lower).

\section{Emitting region}
\label{Emit-Region-Sec}

\subsection{Central absorption}
\label{Central-Abs-Sec}

Marcaide et al. (\cite{Marcaide1995a}, \cite{Marcaide1995b}, \cite{Marcaide1997}, \cite{Marcaide2005b}), 
Bartel et al. (\cite{Bartel2000}, \cite{Bartel2002},
\cite{Bartel2007}), Bietenholz et al. (\cite{Bietenholz2003}, \cite{Bietenholz2005}), 
Alberdi \& Marcaide (\cite{Alberdi2005}), and
Rupen et al. (\cite{Rupen1998}) argued in favor of the radio emission of
SN\,1993J originating in a shell. The determination of the details
of the emitting shell has been difficult. Bietenholz et al. (\cite{Bietenholz2003},
\cite{Bietenholz2005}) and Marcaide et al. (\cite{Marcaide2005a}) suggest that the emission
appears absorbed in the central part compared to the emission
expected from an optically thin shell. Here we present a new data
analysis to show that this is indeed the case.

We estimated $X_{ICP}$, $X_{OCP}$, $Y_{max}$, and $Y_{c}$ and,
from them,  $\rho_{1}$ and  $\rho_{2}$ (see Sect.
\ref{Considerations-Sec}) from simulated shell emissions as well
as from our maps. The shell emission that we used in our
simulations are of two types: (1) emission from an optically thin
shell, and (2) emission from an optically thin shell with a central
absorption that totally blocks the emission from the backside of
the shell out to the shell's inner radius. This blockage could be
due to absorption of the synchrotron radiation by the ionized
ejecta in the line of sight.

\begin{figure}[!t]
\centering
\includegraphics[height=5cm]{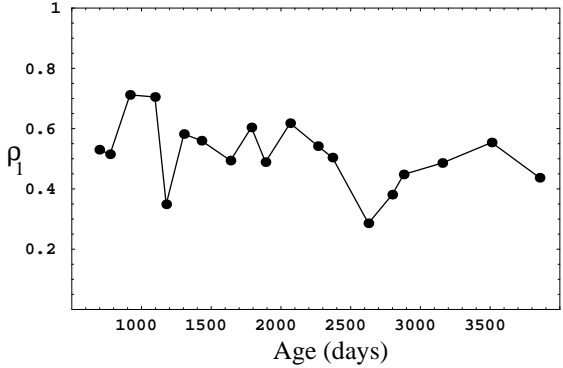}
\caption{Estimates of a measure of the central absorption (see
text) as a function of time.} \label{ratio-CMax}
\end{figure}

\begin{figure}[!t]
\centering
\includegraphics[height=5cm]{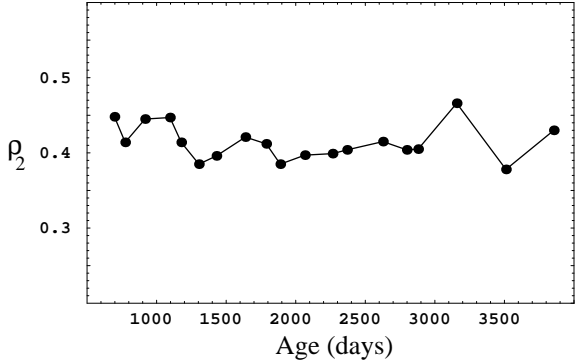}
\caption{Estimates of a measure of the fractional shell width (see
text) as a function of time.} \label{ratio-IO}
\end{figure}

In the simulations, we used three values of the fractional shell
width. The values were centered on the estimate given by
Marcaide et al. (\cite{Marcaide1995a}). For each value, we considered the two types
of emission described above, namely, with and without central
absorption. Table \ref{tab-ratios} shows the ratios $\rho_{1}$ and
$\rho_{2}$ for these simulations. We then determined $\rho_{1}$
and $\rho_{2}$ from our observations. The results are shown in
Figs. \ref{ratio-CMax} and \ref{ratio-IO}. We obtain the
following mean values: $\rho_{1} = 0.51 \pm 0.13$ and $\rho_{2} =
0.41 \pm 0.03$ (all uncertainties quoted in this paper are
standard deviations.)

A conclusion can be readily drawn  from a comparison of these mean
values with the values given in Table \ref{tab-ratios}. The values
of $\rho_{1}$ for the models without absorption and with
absorption in Table \ref{tab-ratios} are included in the ranges
0.75-0.80 and 0.52-0.54, respectively. The observational result
clearly favors the model with absorption and supports the results
published earlier.

For estimating $\rho_{1}$ and $\rho_{2}$, we used only maps
corresponding to a source size larger than twice the beam of the
interferometer to ensure good shell resolution in the maps. For
this reason, we did not use the first 5 epochs of Table
\ref{table-all}.

\begin{table}[!t]
\begin{minipage}[t]{\columnwidth}
\caption{Computed ratios $\rho_{1}$ and $\rho_{2}$ (see text) for
different shell emission models.}
\label{tab-ratios}
\centering
\begin{tabular}{cc|cc}
$\xi$\footnote{Defined in Table \ref{table-bias}.} & 
Absorption\footnote{Defined in Table \ref{table-bias}.} & $\rho_{1}$ & $\rho_{2}$ \\
\hline
0.35 & YES & 0.54 & 0.44 \\
0.30 & YES & 0.52 &  0.46 \\
0.25 & YES & 0.52 & 0.47 \\
\hline
0.35 & NO & 0.80 & 0.37 \\
0.30 & NO & 0.78 & 0.39 \\
0.25 & NO & 0.75 & 0.41 \\
\end{tabular}

\end{minipage}
\end{table}

\subsection{Shell width}
\label{Shell1-Sec}

\begin{figure}[!t]
\centering
\includegraphics[height=5cm]{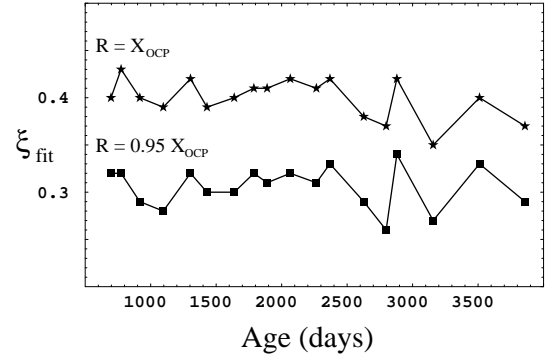}
\caption{Fractional shell widths versus supernova age. The widths are
determined from 6\,cm data by fitting the visibilities as explained
in the text. The source radius is fixed to the value estimated
with the CPM (stars) and to 95\% of this value (filled squares).}
\label{shellwidthtest}
\end{figure}

The shell width can be estimated as explained in Sect.
\ref{Fourier-Sec}. To begin, we fixed the source radius to the
value determined with the CPM, even though we knew that this value
is biased differently depending on the true shell width, as
illustrated by the simulations presented in Table \ref{table-bias}. That
is, we used the value $X_{OCP}$ as a first approximation to the
value R. The estimates obtained for the fractional shell width,
$\xi$, were roughly the same for all epochs as shown in Fig.
\ref{shellwidthtest}. The average estimate is 0.40 $\pm$ 0.04.

However, to obtain a reliable determination of the fractional
shell width we had to know the biases in the determination of the
source radius with the CPM to correct for them and to use the
corrected R values in the model fitting. For instance, a bias of
5\% in the source radius would translate into a decrease in the
fractional shell width from 0.4 to 0.3, as shown in Fig.
\ref{shellwidthtest}.

\begin{figure}[!t]
\centering
\includegraphics[height=5cm]{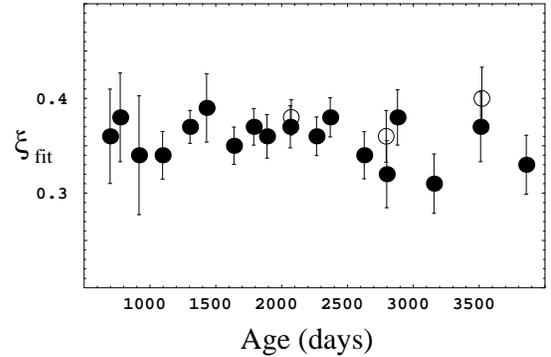}
\caption{Fractional shell widths vs. supernova age. These widths
are determined from 6\,cm data (filled circles) and 18\,cm data (empty
circles) by fitting the visibilities as explained in the text. For
consistency (see the text), the source radius is fixed to 0.975
times the value estimated with the CPM.} \label{shellfinal}
\end{figure}

As we can infer from Fig. \ref{fig52}, estimates of $R_{fit}$
with the fractional shell width fixed at 0.25, 0.30, and 0.35,
yielded unweighted, average values of the ratio $R_{fit}/R_{CPM}$
of 0.93, 0.95, and 0.97, respectively, with an standard deviation
of 0.01 in each case. As mentioned earlier, when we study the bias
in the determination of the source radius with the CPM, using
models of different widths, we obtain the results shown in Table
\ref{table-bias}. By requiring consistency, we find from Table
\ref{table-bias} the bias in the model corresponding to the ratio
$R_{fit}/R_{CPM}$ determined for that same model ($R$ and
$X_{OCP}$ in Table \ref{table-bias} correspond to $R_{fit}$ and
$R_{CPM}$, respectively). This consistency in the model with
absorption is obtained only for a fractional shell width of 0.35,
which yields a 2.5\% bias and a ratio $R_{fit}/R_{CPM}$ of
$0.97\pm0.01$. For the models with shell widths 0.30 and 0.25, the
corresponding values are 3\% and $0.95\pm0.01$, and 4\% and
$0.93\pm0.01$, respectively. These pairs of values are not as
consistent as the pair corresponding to the value 0.35 for the
fractional shell width.

We took another redundancy step in this testing: we generated
visibility data for conditions similar to the observational ones
using a fractional shell width of 0.35. Then we executed the previous
procedures to determine the values of the ratios $R_{fit}/R_{CPM}$
using models with fractional shell widths of 0.25, 0.30, and 0.35.
The values obtained for the ratios $R_{fit}/R_{CPM}$ were 0.94,
0.96, and 0.98, respectively. In all cases the uncertainty was
about 0.01. These estimates are very close to the values of the
ratios that we obtained using the real observations and those same
models.

The conclusion seems inescapable: the fractional shell width of a
model (with total absorption of emission from the shell region
behind the ejecta) most compatible with the data is 0.35, with the
estimate of the outer radius of the shell being about 0.975 times
the estimate provided by the CPM. As a final refinement to our
determination of the fractional shell width, we re-estimated $\xi$
for all 6~cm data using as a source radius 0.975 times the CPM
estimate. The results are shown in Fig. \ref{shellfinal}. The
weighted mean of these estimates of the fractional width is
$0.359\pm0.019$. The corresponding results that we obtained for the
18~cm data are also shown in Fig. \ref{shellfinal}. In this case, the weighted
mean is $0.378\pm0.013$, implying that the fractional
width might be slightly larger at 18~cm than at 6~cm.

\section{Discussion of results}
\label{Discussion-Sec}

\subsection{Supernova expansion}

In the previous sections, we presented the analysis
of our data. To determine the characteristics of the expansion, we
estimated the source size for each epoch using two methods: (1)
the CPM described in Sect. \ref{CPM-Sec}, which uses the
map of the source in the analysis; and (2) a fit of a model directly
to the visibilities. The results obtained by the two methods are
consistent. However, the CPM is more precise than the fit of a
model to the visibilities, and in the latter, one must use an
a priori model. In any case, the results are consistent to within
1\% if the bias between the two determinations is taken into
account. The determination of this bias, and its constancy in
time, is shown in Fig. \ref{fig52}. Thus, the 
measured expansion rate is the same for both methods. The results, shown in
Fig. \ref{fig51} for the 6\,cm data, are from the use of the CPM.
As already discussed in Sect. \ref{Expan1-Sec}, a break in rate
at about day 1500 after explosion can be readily seen in Fig.
\ref{fig51}. The expansion index $m$ before that break takes the
value $0.845\pm0.005$ and after the break $0.788\pm0.015$.
Remarkably, the 18~cm data do not follow the 6~cm data after the
break. Rather, the former seem to fall where one would expect for
a prediction based on the value of the index before the break.

The 18\,cm data depart significantly from the 6\,cm data. The
difference of the expansion indices is $0.057 \pm 0.016$, that is,
the ratio of the source size at 18\,cm to the size at 6\,cm evolves
in time as $t^{0.057 \pm 0.016}$. A physical model of the source
emission should explain this evolution. We propose models for
which the 18\,cm data systematically depart from the 6\,cm data.
Then, we discuss our results for the shell width.

A straightforward interpretation of these unexpected expansion
results states that at the longer wavelength the emitting region
extends to the outer shock front in the mini-shell model, while at
the shorter wavelength the emitting region is progressively
radially smaller and therefore appears to grow at a slower rate
than the radius of the outer shock front, this effect becoming
discernible after a given epoch, in our case about 1500 days after
explosion. In other words, the size of the emitting region should
be wavelength dependent. We have attempted to physically model
this wavelength dependence; in the process, we eliminated one 
possible physical explanation but identified two other promising 
ones.

\subsubsection{Synchrotron mean-life of electrons}
\label{synchmeanlife}

We excluded an explanation based on the mean-life of the emitting
electrons. In principle, if electron acceleration occurs near
the contact discontinuity, the electrons that emit at 18\,cm
(1.7\,GHz) should travel further out than the particles that emit at
6\,cm (5\,GHz) since their mean-life should be longer. The problem
with this explanation is that the mean-life of all of those
relativistic electrons is far too long. Using the equations 3.28
and 3.32 from Pacholczyk (1970), we estimate a mean-life of 19 yr
for a critical frequency $\nu_c$ = 5\,GHz (6\,cm wavelength) and a
magnetic field $B$ = 0.1 Gauss (corresponding to a supernova age
of 1500 days according to P\'erez-Torres et al. \cite{PerezTorres2001}). This
mean-life is far too long to be compatible with the time needed
for the electrons to traverse the shocked circumstellar region,
even for random trajectories. Fransson \& Bj\"ornsson (\cite{Fransson1998})
proposed a similar mechanism to obtain a wavelength-dependent
emitting region. It was based on the mean-life of the emitting
electrons, but they assumed that the emission would be generated
in the neighborhood of the forward shock. Their mechanism can be
discarded on the same grounds as was ours.

\subsubsection{Radially-decreasing magnetic field in the shell}

At present, there is no strong theoretical justification to
consider a magnetic field dependence on distance from the constant
discontinuity, inside the supernova shell. However, this dependence
on distance is plausible, because the field amplification might
take place in the turbulent regime next to the contact
discontinuity (e.g., Chevalier \& Blondin \cite{Chevalier1995}). With this
motivation, we consider, for example, a linear decrease in the
magnetic field with distance from the contact discontinuity and
find the observational consequences. Our conclusions will not
qualitatively depend on the particular shape of this decrease.
Thus, we assume for the magnetic field the expression

\begin{equation}
B(D) = B_{c}\times\left(1-\frac{D}{D_{max}}\right) \label{linB}
\end{equation}

\noindent where $D$ is the distance from the contact discontinuity,
$D_{max}$ is the maximum range of the field, which cannot exceed
the distance to the position of the forward shock, and $B_{c}$ is
the magnetic field at the contact discontinuity, which has been
chosen such that the average magnetic field over the emitting
region is 0.057 Gauss for day 3200, as suggested by P\'erez-Torres
et al. (\cite{PerezTorres2002}).

We consider two models, which have in common the essential
ingredient of a radially decreasing magnetic field. The first one
concerns synchrotron aging of the emitting electrons and the
second the finite sensitivity of the interferometers. We describe
each of them in turn.

Synchrotron aging translates into a deviation of the electron
energy distribution from the canonical $N \propto E^{-p}$
distribution at high energies. Chandra, Ray, \& Bhatnagar (\cite{Chandra2004})
suggested synchrotron aging in SN\,1993J, based on their observed
radio spectrum. We should note however that this suggestion is not
supported by the work of Weiler et al. (\cite{Weiler2007}).

\begin{figure}[t]
\centering
\includegraphics[height=4cm]{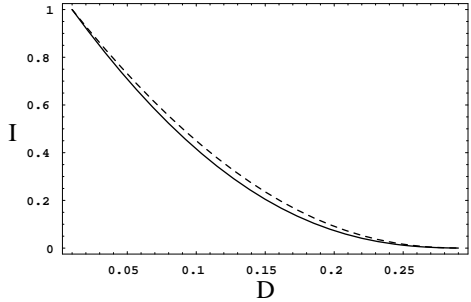}
\caption{Radial emission intensity profiles at 6\,cm (solid line)
and 18\,cm (dashed line), using a linear radial decrease in the
magnetic field and a synchrotron aged electron energy distribution
(see text). Each of the profiles is normalized to its
corresponding emission at the contact discontinuity. $D$ is the
distance from the contact discontinuity in units of the source
radius.} \label{fig-theor-intens}
\end{figure}

\begin{figure}[t]
\centering
\includegraphics[height=4cm]{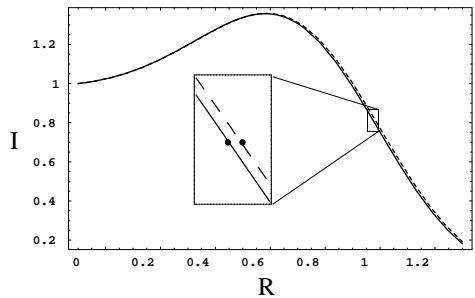}
\caption{Profiles of the azimuthal averages of the maps obtained
from the radial intensity distributions shown in Fig.
\ref{fig-theor-intens} normalized to the corresponding intensity
at the source center. The units for the x-axis are normalized to
the source size. As in the previous figure, the solid line
corresponds to 6\,cm and the dashed line to 18\,cm. The dots shown
in the enlargement correspond to the outer common point of each
profile.} \label{fig-azimaver}
\end{figure}

If we assume emission within an optically thin medium, we can
compute for a synchrotron-aged electron population (see Appendix
\ref{Ageing-App}) the 2D image corresponding to the emission
profile in Fig. \ref{fig-theor-intens} for each of the radio
wavelengths. For each wavelength, the profile corresponding to the
azimuthal average of the 2D image, convolved with a Gaussian beam,
is shown in Fig. \ref{fig-azimaver}. We apply the CPM (see
Sect. \ref{CPM-Sec}) to estimate the size of each image.

Clearly, as in Fig. \ref{fig-theor-intens}, the profile reaches
a given level of intensity further out at the longer wavelength,
thus increasing the size estimate of the CPM by about
0.5\% over the corresponding estimate at the shorter wavelength.
This increase falls short of the observational result by a factor
of 4, but it does go in the right direction. A steeper radial drop
in the magnetic field would decrease the shortfall. For example,
an exponential drop would decrease it by about a factor of 2.
A possible high-energy cutoff in the relativistic electron
distribution would also contribute in the same direction to yield a
size estimate larger at 18\,cm  than at 6\,cm.

A consequence of the previous explanation is that there should
also be  a difference in the size estimations between 6 and
3.6\,cm, although the difference should be smaller than that
between 18 and 6\,cm (see Eq. \ref{intensityratio} in Appendix
\ref{Ageing-App}). We do not find such a difference in our own
data. However, our data at 3.6\,cm are restricted to the earliest
epochs and do not overlap temporally with the data at 6\,cm.

\begin{figure}[t]
\centering
\includegraphics[width=6cm]{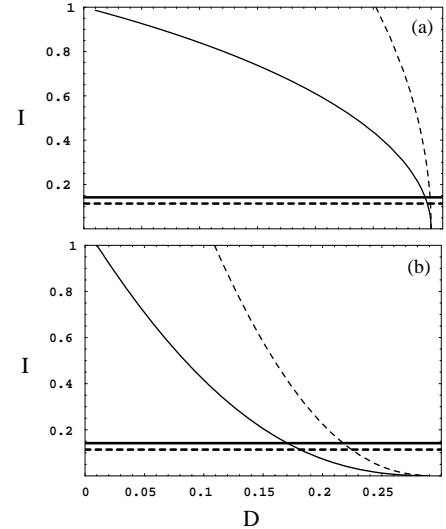}
\caption{Radial emission intensity profiles at 6\,cm (solid line)
and 18\,cm (dashed line) normalized to the emission at 6\,cm at the
contact discontinuity. The horizontal lines (6\,cm, thick solid
line; 18\,cm, thick dashed line) indicate realistic noise levels:
(a) for a constant magnetic field in the emitting region; and (b) 
for a linear radial decrease in the magnetic field (see Eq.
\ref{linB}). A synchrotron-aged electron energy distribution (see
Eq. \ref{chandrapop}) was used in both cases. D is the
distance from the contact discontinuity in units of the source
radius.} \label{threshold-effect}
\end{figure}

The limited sensitivity of an interferometric array enhances considerably
the difference of the source size at 6 and 18\,cm, if
measurable. Indeed, as shown in Fig. \ref{threshold-effect}(b),
which is an enlargement of the outer region of Fig.
\ref{fig-theor-intens} but where both 
types of emission are normalized to
the 6\,cm emission at the contact discontinuity, the intersection
of these curves with a realistic noise level (i.e., obtained in
the simulation using typical antenna system temperatures) takes
place at a quite different radial position at 6\,cm than at 18\,cm
for a linear radial decrease in the magnetic field, but not for a
model with a constant magnetic field (Fig.
\ref{threshold-effect}(a)). Both intersections also occur at
smaller values of $D$ for Fig. \ref{threshold-effect}(b) than
for Fig. \ref{threshold-effect}(a).

As indicated in Fig. \ref{threshold-effect}(a), for a constant
magnetic field the realistic noise intersection for both
wavelengths occurs at practically the same value of D. Hence, the
measurement of the source radii in the maps corresponding to those
source emission intensity profiles would be practically unaffected
by the noise. (Note that the profiles at 6 and 18\,cm shown in
Fig. \ref{threshold-effect}(b) are the same as in Fig.
\ref{fig-theor-intens}, but appear to differ only because the
emission at both wavelengths was normalized to the 6\,cm
emission level. For the model considered, the source emission is
stronger at 18\,cm than at 6\,cm.) For unrealistically large noise
levels (i.e., a large fraction of the source flux density per
beam), a difference in the sizes at the two wavelengths would in
principle also be noticeable, but in practice at those noise
levels we would not even be able to reconstruct the VLBI maps with
sufficient quality to detect the effect we discuss here.

For a magnetic field that decreases radially (Fig.
\ref{threshold-effect}(b)), the source emission above the noise
extends to a larger D at 18\,cm than at 6\,cm. Therefore, the source
radius would also be smaller at 6\,cm than at 18\,cm. The difference
in the size estimates between 6 and 18\,cm at day 3200 (see
previous subsection) is about 2\% and has the ``right''  sign.

\subsubsection{Changes in the opacity of the ejecta}

The effects considered in the previous sections may account for
some of the differences in the expansions observed at 6 and 18~cm.
However, these effects seem to be insufficient to account for a
4\% difference in the sizes at day 3200 or for a larger difference
at later epochs. On the other hand, flux density measurements,
made with the VLA at epochs where we have data 
at both 6 and 18~cm,
also show a trend that is worthwhile discussing. Fransson \&
Bj\"ornsson (\cite{Fransson1998}) suggested that for late epochs (beyond day 1000
after explosion) the spectral index between 6 and 18~cm should
remain rather constant, $\alpha \sim 0.75$ ($S_\nu \sim
\nu^{-\alpha}$). However, observationally the spectral index of
the supernova does not remain constant but decreases with time
(see our Table \ref{VLAfluxes} and Fig. 3a of Weiler et al.
\cite{Weiler2007}). How can we explain this change?

\begin{table}[!t]
\begin{minipage}[t]{\columnwidth}
\caption{Spectral indices, $\alpha$, determined from our VLA data
for a subset of observations for which we have quasi-simultaneous
6 and 18~cm observations (M\,81 used as calibrator).}
\label{VLBIfluxes}
\centering
\begin{tabular}{c|c}
Age (days) &$\alpha$\\
   &  \\
\hline
2794  & 0.68$\pm$0.01 \\
2820\footnote{From P\'erez-Torres et al. (\cite{PerezTorres2002}).} 
&  0.67$\pm$0.02 \\
3511 & 0.555$\pm$0.010 \\
3858 & 0.50$\pm$0.04 \\
\end{tabular}
\end{minipage}
\label{VLAfluxes}
\end{table}

\begin{figure}[t]
\centering
\includegraphics[height=5cm]{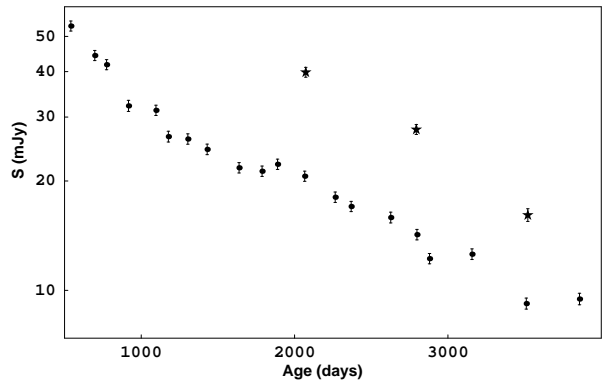}
\caption{Time evolution in the total map flux densities of
SN\,1993J, as determined from the VLBI data. See Table
\ref{table-all} for the values. The standard deviations
were estimated by adding 3 times the rms map noise to 3\% of the total 
flux density, to account for calibration systematics. Circles and stars 
correspond to 6 and 18\,cm data, respectively. Weiler et al. (\cite{Weiler2007}) 
presented similar light curves observed with the VLA. The bump is also very 
conspicuous in their curves at several wavelengths.}
\end{figure}

In Fig. \ref{VLBIfluxes}, we plot the 6 and 18\,cm flux densities
obtained from our VLBI maps and given in Table \ref{table-all} 
(M\,81 was used as flux density calibrator; our estimates may thus 
differ systematically from those of Weiler et al. \cite{Weiler2007}). 
We notice a sharp change in the evolution of the 6\,cm map flux
density after day 1500, which correlates with the change in slope
of the expansion measured at that same wavelength. This
correlation may have some significance. The evolution in the 6\,cm
flux density in Fig. \ref{VLBIfluxes} appears to exhibit an
increase with respect to the evolution expected from previous
epochs. A natural way of obtaining this increase could be to start
receiving emission from a region of the shell that was suppressed
at previous epochs, namely, the emission from the side of the
shell behind the ejecta. An opacity of the ejecta that decreases 
with time making them more transparent to 6\,cm than to 18\,cm
radiation is sufficient.

\begin{figure}[t]
\centering
\includegraphics[height=5cm]{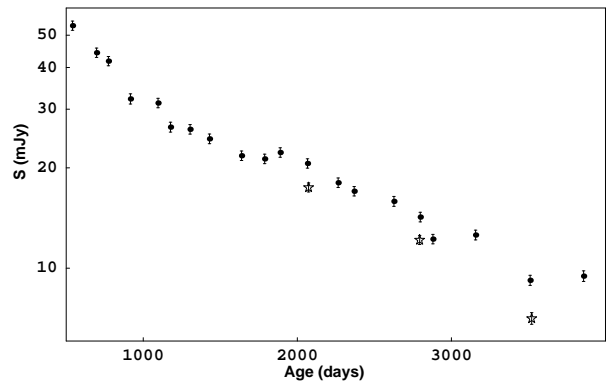}
\caption{Same as previous figure, but with the 18~cm data
converted (empty stars) to a 6\,cm equivalent total 
flux density using a spectral index of value 0.75.} 
\label{VLBIfluxes-18scaled}
\end{figure}

In spite of what has been said above, we transform the 18\,cm flux
densities into ``hypothetical 6\,cm flux densities'' using the
spectral index suggested by Fransson \& Bj\"ornsson (\cite{Fransson1998}) 
and compare them with the true 6\,cm flux densities. Surprisingly,
as shown in Fig. \ref{VLBIfluxes-18scaled}, transformed 18\,cm
flux densities follow the 6~cm flux-density evolution expected
from epochs earlier than day 1500. Hence, a possible
interpretation might be the following: the true spectral index is
$\sim0.75$, as suggested by Fransson \& Bj\"ornsson (\cite{Fransson1998}), or
close to it (Weiler et al. \cite{Weiler2007}), and the spectral index evolution
we observed is due solely to an increase in detected emission at
6\,cm from the shell region behind the ejecta because the latter
becomes more transparent at 6\,cm than at 18\,cm after day 1500.

As noticed in an earlier section and quantified in Table
\ref{table-bias}, a change in the emission-absorption model will also
result in a different estimate of the source size for a given set
of data. Thus, as shown in Table \ref{table-bias}, for a given data set
a model with absorption will yield a larger estimate of the source
size than a model without absorption. The difference in the
estimates can be as large as 2.5\% for a change in opacity from
total absorption to no absorption at all. For a shell model (with
a shell of 0.30 fractional width) of radius $R$ with total
absorption, the size determined by the CPM would be $1.03R$. For
the same model of radius $R$ without absorption, the size
determined by the CPM would instead be only $1.005R$.

Thus, a transition from a model with absorption to one with
no-absorption would result in a decrease in the size estimated, if
either the CPM or fitting to the visibilities is used. Were the
transition (decrease in opacity) to take place over a period of
time, the net effect would be a progressive decrease in size 
estimates during this period, resulting in turn in a decrease in
the estimate of the deceleration parameter (namely, the estimated
$m$ would be smaller than the true $m$ because of the 
decrease in the absorption in the source). Consequently, after the ejecta become
fully transparent to the 6~cm radiation, the true deceleration
parameter $m$ will be recovered. Unfortunately, given the flux
density evolution in the source (Weiler et al. \cite{Weiler2007}), this
recovery will be unlikely to take place while SN\,1993J can still
be mapped at 6\,cm.

\subsection{Shell width}

We determined the shell width as a fraction of source
radius, using for the latter the bias-corrected results from the
CPM and model fitting to the visibilities as explained in Sect.
\ref{Shell1-Sec}. The results given in Fig. \ref{shellfinal}
show a rather similar fractional width determination for all
epochs. The average values of the 6 and 18\,cm results infer
fractional shell width values of $0.359\pm0.019$ and
$0.378\pm0.013$, respectively.

These previous estimates were based on a model that assumed total
absorption of the emission from the back of the shell (behind the
ejecta). Partial absorption is clearly also compatible with our
data and would yield a smaller fractional value for the shell
width. In this sense, the result given above is an upper bound to
the fractional shell width.

We note that the uncertainties (and scatter) in the shell width
estimates from the 6\,cm data for the supernova age range of
1500-2500 days, as shown in Fig. \ref{shellfinal}, are generally
smaller than for earlier and later ages. For earlier epochs, the
supernova size is smaller and the determination more difficult,
whereas for later epochs the flux density of the source is lower
and hence the data are relatively noisier. The 6\,cm data for the
range 1500-2500 days (corresponding to 6 epochs of observation)
are optimal, in a sense, for the determination of the shell width.
Thus, we used these data to simultaneously determine the
source size, the shell width, and the degree of absorption of the
model. These data suffice to reliably obtain the estimates of all
3 parameters. The results are shown in Table \ref{fit-of-all}. The
average estimate of the fractional shell width is $0.31\pm0.04$
and the degree of absorption in the corresponding model is
$(80\pm14)$\%. The large uncertainties are caused mostly by the 
first two epochs. [Using only the last four epochs, we obtain
$0.33\pm0.04$ and $(73\pm7)$\%, respectively.] The ratio
$R_{fit}/X_{OCP}$ is $0.97\pm0.03$, which is consistent with the
theoretical bias expected for a model with 80\% absorption and a
fractional shell width of 30\%: $0.968$. As expected, less
absorption translates into a smaller shell width estimate.

\begin{table*}[!t]
\begin{minipage}[t]{17cm}
\caption{Model fitting results for epochs between 1500 and 2500 days.}
\label{fit-of-all}
\centering
\begin{tabular}{c|ccc}
Age & $R_{fit}/X_{OCP}$~\footnote{Fitted source radius, normalized to the 
CPM estimate.} 
&$\xi_{fit}$\footnote{Fitted fractional shell width.} 
&Absorption\footnote{Fitted percentage of absorption by the ejecta.} \\
(days) & & & (\%) \\
\hline
1638 & 0.935$\pm$0.016 & 0.24$\pm$0.03 & 108$\pm$4 \\
1788 & 0.952$\pm$0.010 & 0.294$\pm$0.017 &  91$\pm$5 \\
1889 & 1.009$\pm$0.013 & 0.374$\pm$0.013 & 77$\pm$2 \\
2066 & 0.987$\pm$0.013 & 0.315$\pm$0.019 & 61$\pm$5 \\
2265 & 1.000$\pm$0.012 & 0.355$\pm$0.017 & 80$\pm$6 \\
2369 & 0.960$\pm$0.012 & 0.27$\pm$0.02 & 67$\pm$6 \\
\end{tabular}
\end{minipage}
\end{table*}

\begin{figure}[!t]
\centering
\includegraphics[height=5cm]{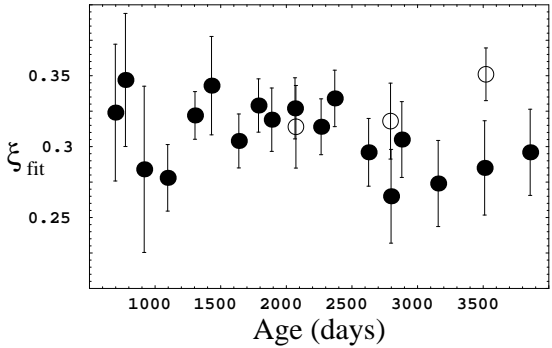}
\caption{Relative shell widths versus supernova age for a model 
with 80\% absorption. The widths are determined from 6~cm
(filled circles) and 18\,cm data (empty circles) by model-fitting
the visibilities. For consistency (see text), the source radius is
fixed at 0.968 times the value estimated with the CPM.}
\label{shellfinal80}
\end{figure}

In view of the previous results about absorption, we re-estimated the
fractional shell width using data from all epochs, as before, but
now for a model with a fixed absorption of 80\%. After taking into
account the corresponding bias of the CPM, 0.968, we present the
results in Fig. \ref{shellfinal80}. This figure contains our
most accurate estimates of the fractional shell widths. The
average values for the 6\,cm and 18\,cm results are $0.31\pm0.02$ and
$0.335\pm0.017$, respectively. The degree of absorption at 18\,cm
cannot be determined from our data; it might well be 100\%, in
which case the estimate of the fractional shell width at this
wavelength would be $0.378\pm0.013$.

Figure \ref{shellfinal80} hints at a possible decrease in the
fractional shell width at 6\,cm for a supernova age beyond day
2500, while the corresponding estimates for 18\,cm seem to show, if
anything, the opposite trend. We expect this kind of evolution for
each of the 3 mechanisms proposed in the previous section to
explain the characteristics of the expansion at 6\,cm:

\begin{itemize}

\item (1) Synchrotron aging would reduce the extent of the emitting
region and hence the source size (outer radius of the emitting
region). Given that the inner radius would be unchanged by
this mechanism, the fractional shell width would decrease. This
mechanism would account for about a 0.5-1\% decrease in the source
size (1.5-3\% in the shell width).

\item (2) A decreasing magnetic field with radial distance, coupled with
the limited sensitivity of the interferometers would account for a
decrease in the source size and, hence, by the same argument as 
for mechanism (1), of the shell width. This mechanism would 
account for about a 2\% decrease in the
source size (6\% in the shell width). Mechanism (1) would 
correspond to a real decrease, mechanism (2) to a decrease caused by an instrumental
effect. In both cases, a radially decreasing magnetic field is
necessary.

\item (3) A changing opacity translates into changing source size
estimates when a fixed source model is used in the size estimation
process. Part of the enhanced deceleration beyond day 1500,
apparently seen at 6\,cm, might be caused by use of a model with
fixed absorption. Decreasing absorption would thus lead to
underestimates of the source size. Since in fitting the shell
width, the source size is a fixed parameter, an underestimate of
the source size would translate into an underestimate of the shell
width. This mechanism would account for a maximum of 2.5\% (change
from totally opaque to totally transparent ejecta at 6\,cm) in the
source size (up to 7.5\% of the shell width). However, since the
absorption is apparently not lower than about 80\%, this
mechanism could account for only a fraction of the previous
estimates.

\end{itemize}

\section{Comparison with other VLBI results}

\subsection{Supernova expansion}

Bartel et al. (\cite{Bartel2002}) reported VLBI observations of SN\,1993J made
at several wavelengths over a similar time range as we now report.
Their analysis of the VLBI data was carried out with different
tools than ours. They estimated the source size by fitting models
directly to the visibilities using a modified version of task
UVFIT of the NRAO AIPS Package. The model used in those fits was a
three-dimensional spherical shell of uniform volume emissivity and
fractional shell width of 0.2. They estimated the outer radius of
the shell and also the plane-of-the-sky coordinates of its center
with respect to the phase center of the supernova. In a previous
paper, Bietenholz et al. (\cite{Bietenholz2001}) concluded that these center coordinates
remain fixed (within the uncertainties) with respect to the core
of M\,81; thus, it is unclear to us why they were included as free
parameters (presumably as a check on their previous work).
Including these parameters is equivalent to determining the slopes
of the imaginary part of the visibilities. Because of the remarkable
circular symmetry of the source, these imaginary parts contain no
significant structure information, especially those for the short
baselines that are the most relevant to the fits.

As explained earlier in this paper, we determined the shell outer
radius with two methods. For each, we assumed the same center for
all epochs and found that, to within the noise of the
measurements, the imaginary part of the visibilities vanished and
so we could fit the outer shell radius using only the real part.
As a consequence, we estimated fewer parameters in the fit and,
perhaps for that reason, the solutions are more stable in our case
than in Bartel et al. (\cite{Bartel2002}). We also used three-dimensional
spherical shells of uniform emissivity, but including absorption
by the supernova ejecta. Hence, Bartel et al.'s results are not
directly comparable to ours.

Bartel et al. (\cite{Bartel2002}) determined a deceleration parameter for each
of four supernova age ranges (see their Table 4), noting that
``the four time intervals with different decelerations can be
distinguished by eye.'' For each of the four age ranges, Bartel et
al. determined a deceleration for each observed wavelength and
noticed differences in the determinations. They also considered
decelerations obtained from the combined data set as being
representative.

Based on their determination of the deceleration for each of the
four supernova age ranges, Bartel et al. inferred an increasing
deceleration of the expansion followed at about day 1893 by a
slowing deceleration. This evolution in the deceleration, they
stated, provides support to the predictions made by Mioduszewski
et al. (\cite{Mioduszewski2001}) from their hydrodynamic simulations. Those
simulations depend very strongly on certain features of the
density profile of the ejecta (explosion model 4H47 provided by
Nomoto's group), which are of some concern. Simulations with less
specific -and more accepted- explosion models would not have
produced these results. Hence, a re-analysis of the expansion data
provided by Bartel et al. (\cite{Bartel2002}) might be of interest. A
re-analysis of all VLBI data from Bartel's group was part of the
Ph.D. thesis of Mart\'i-Vidal (2008) and will be published elsewhere.

We fit the expansion model that we described in Sect.
\ref{Expan1-Sec} to the data published by Bartel et al (\cite{Bartel2002}).
This expansion model considers two deceleration regimes and the
time of change between them, yielding three parameters to be
estimated. At 6\,cm, we used 20 outer-shell radius determinations
from Bartel et al. (\cite{Bartel2002}), which ranged from day 223 through day
2996, and obtained the deceleration parameter estimates of $m_{1} =
0.85 \pm 0.03$ and $m_{2} = 0.78 \pm 0.04$, with a break at day
$900 \pm 500$. At 3.6\,cm, the 26 data points, which extend from day
50 to day 2787, yield the estimates $m_{1} = 0.96 \pm 0.04$
and $m_{2} = 0.80 \pm 0.02$ with a break at day $320\pm90$.
However, if we ignore the 3.6\,cm data from before day 223, we
obtain $m_{1} = 0.84 \pm 0.04$ and $m_{2} = 0.78 \pm 0.04$ with
the break at day $700\pm600$. In all cases, the uncertainties
given are scaled such that the reduced chi-squares of the fits are
unity. Given the uncertainties shown, the parameter estimates for
the 6 and 3.6\,cm data for days after 223 are remarkably similar.

These estimates of the deceleration parameters are also very
similar to those we obtained from our own data using what we
think is a more accurate method, the CPM: $0.845\pm0.005$ and
$0.788\pm0.015$, although we find the break time to be at day
$1500\pm300$ in our case. Thus, we conclude that the Bartel et al.
data are compatible with just one change in deceleration and that
there is no need to invoke changes in deceleration at other times.
The agreement of the results given by Bartel et al. with the
hydrodynamic model of Mioduszewski et al. may therefore not be
significant.

Bartel et al. (\cite{Bartel2002}) also published estimates of the source
size for 8 epochs at 13\,cm and for 3 epochs at 18\,cm. The 13\,cm
estimates are consistent with those for 3.6 and 6\,cm for the first
1000 days but for later epochs the estimates are systematically
larger than for 6\,cm. Except for their estimate around day 1000,
those results are consistent with our findings, to within the
estimated standard errors. However, the 13\,cm data do not show a
clear trend of departure from the 6\,cm curve. The Bartel et al.
18\,cm estimates are also larger than the 6\,cm estimates for the
same epochs, but have significant scatter. We reanalyzed the VLBI
observations from Bartel's group for days 1692 (6\,cm) and 3164
(18\,cm). These size determinations are consistent with our
determinations from our own data and indicate that Bartel et al.'s
estimate at 18\,cm is an underestimate ($\sim 4$\%). In any case,
since the 18\,cm estimates are larger than those at 6\,cm at
the latest epochs, those authors (see Fig. 6 in Bartel et al.
\cite{Bartel2002}) interpreted this as slowing deceleration. Even the 
last 6\,cm estimates of Bartel et al. are probably overestimates, if the
true shell width decreases in that period, as we concluded in
Sect. \ref{Discussion-Sec}. Keeping a model with a fixed
fractional shell width, with no absorption, in fitting data
corresponding to a decreasing shell width, or monotonically
decreasing absorption by the ejecta, leads to a progressively
increasing overestimate of the size. This overestimation of the
sizes from the 6\,cm data has two consequences: it reinforces one's
impression of a slowing deceleration and prevents one from
discerning the progressively increasing difference in the sizes at
6 and 18\,cm.

\subsection{Shell width}

Bartel et al. (\cite{Bartel2002}) required their data to meet a set of criteria
before using them to estimate the shell width. A total of 16
epochs  fulfilled their criteria (10 at 3.6\,cm and and 6 at 6\,cm).
They estimated a fractional shell width of $0.25\pm0.02$ by
fitting a spherical shell model, without any absorption, to their
data. They deferred to a later article by Bietenholz et al.
(\cite{Bietenholz2003}) discussion of adding (possible) absorption in their
model. In the latter paper, the authors estimated the fractional
shell width as $0.25\pm0.03$, using a simple disk model to
simulate a 25\% absorption of the radiation from the central part
of the source. With such a small amount of absorption, the estimate of the
fractional shell width is about the same as with a model without
absorption. They mentioned that using a larger disk model and a
stronger absorption the relative shell width could be as large as
0.35, but they excluded this option because they considered the
fit to the data to be worse. In our opinion, the disk model they used to
simulate the absorption was too restrictive and directly caused
their inadequate fit (see their Fig. 13).

\section{Comparison with results from optical observations}

Fransson et al. (\cite{Fransson2005}) presented HST ultraviolet spectra from the
nebular phase of the expansion of SN\,1993J (days 670-2585). They
found that the spectrum remained remarkably constant in time.
However, they identified small temporal changes in the shape of the
Mg~II line, which, as mentioned by these authors, changes in
concordance with the shape of the H$_{\alpha}$ line observed by
Matheson et al. (\cite{Matheson2000a}). Since the changes between days 1063 and
1399 are small, Fransson et al. (\cite{Fransson2005}) averaged the spectra from
those epochs to increase the signal-to-noise ratio, and fit these
data and the H$_{\alpha}$ line data on day 976 to a model with a constant
emissivity shell of inner velocity V$_{in}$ and outer velocity
V$_{out}$. As shown in their Fig. 3, a combination of
V$_{in}$=7\,000 and V$_{out}$=10\,000\,km\,s$^{-1}$ fits the boxlike
data rather well. As good a fit can be obtained either with a
rather thin shell, regardless of the emissivity structure, or with
a thicker shell of constant emissivity. There are also 
indications
that V$_{in}$=6\,000\,km\,s$^{-1}$ could also fit the lines (see the discussion in
Fransson et al. \cite{Fransson2005}). Thus, one should take the 7\,000~km~s$^{-1}$
as an upper bound to the lowest velocities in the optically
emitting shell. For an homologous expansion, these velocity
measurements would translate into a 30\% wide shell.

How does this optically emitting shell relate to the radio
emitting shell? For the same epochs, it is rather remarkable that
the upper and lower velocities in the optical and radio shells are
nearly the same. By estimating the velocity of the outer
radio surface from the expansion shown in Fig. \ref{fig51}, and
assuming the distance of $3.63\pm0.34$\,Mpc to M\,81 (Freedman et
al. \cite{Freedman1994}) as the distance to SN\,1993J, we derive the range
10\,000$-$10\,500~km~s$^{-1}$ for days 1063-1399, which implies
velocities in the range 6\,900$-$7\,250\,km\,s$^{-1}$ in the inner radio
surface, for our determination of the fractional shell width,
which remains nearly constant at $0.31\pm0.04$ (i.e., the shell
expands self-similarly). The optical emission, instead, is thought
to originate in a cool dense shell in the shocked ejecta (e.g.,
Fransson \cite{Fransson1984}). This shell is necessarily spatially thin.
Consequently, the optically emitting shell cannot have an
homologous structure. The problem with this optical-radio
comparison is that the inner part of the radio emitting shell,
next to the contact discontinuity, is further from the center than
is the optically emitting shell, but at lower velocities than most
of the shell material emitting in the optical. Were these optical
velocities to be expansion velocities, would the optical shell
eventually enter the radio shell?

\begin{figure}[!h]
\centering
\includegraphics[width=7.5cm]{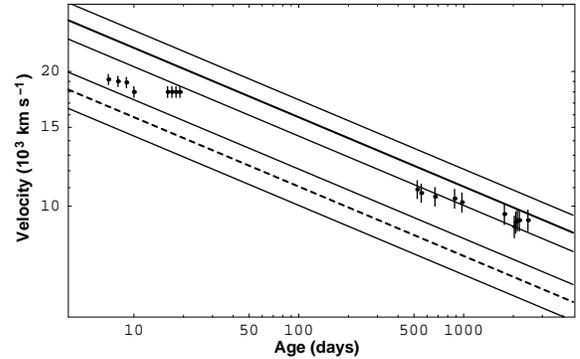}
\caption{Velocities of the inner (dashed line) and outer (solid
line) radio shell surfaces computed from our expansion model using
$m=0.845$ and a distance of $3.63\pm0.34$\,Mpc to SN\,1993J. The
thin lines at each side of the dashed and solid lines indicate the
uncertainties. Filled circles (and error bars) are the maximum
velocities of the H$\alpha$ line (and their standard deviations)
reported by various authors (see text).}

\label{OptVel}
\end{figure}

Only for the optical emission taking place at the contact
discontinuity would the velocity of the optical lines be equal to
the expansion velocity of the contact discontinuity. In that case,
the optical velocities and the velocities inferred from VLBI would
be directly comparable. In Fig. \ref{OptVel}, we show the
velocities of the inner and outer radio shell surfaces (each with
the corresponding uncertainty) as computed from our expansion
model using $m=0.845$. In that figure, we also plot the maximum
velocities at the blue edge of the H$\alpha$ line
reported by Trammell et al. (\cite{Trammell1993}), Lewis et al. (\cite{Lewis1994}), Finn et
al. (\cite{Finn1995}), and Matheson et al. (\cite{Matheson2000b}), all as given in Bartel et
al. (\cite{Bartel2007}). It is interesting to see that optical velocities
appear closer to the inner shell surface velocity for early days
and further from it at late days, thus indicating a lower
deceleration of the regions responsible for the optical emission
than that of the radio shell. In other words, if this high
velocity optical emission takes place at the tips of the
Rayleigh-Taylor fingers (a kind of effective contact
discontinuity) the comparison of the velocities is indicating a
progressive penetration of the fingers into the shocked
circumstellar medium. Bartel et al. (\cite{Bartel2007}) also arrived at a
similar conclusion. Chevalier \& Blondin (\cite{Chevalier1995}) proposed 
this kind of evolution in the Rayleigh-Taylor fingers from hydrodynamical 
simulations.

\section{Conclusions}

We have studied the growth of the shell-like radio structure of
supernova SN\,1993J in M\,81 from September 1993 to October
2003 with very-long-baseline interferometry (VLBI) observations at
the wavelengths of 3.6, 6, and 18\,cm. We used two methods to
analyze our data: a new method, named the Common Point Method
(CPM), described in detail in Sect. \ref{CPM-Sec} and Appendix
\ref{CPM-App} of this paper; and by fitting a specific parameterized
model to the visibilities, as described in Appendix
\ref{Modelfit-App}. The CPM allows us to accurately estimate the
source size on the sky plane for a (nearly) circularly symmetric
compact radio structure such as SN\,1993J. The model fitting estimate
instead depends strongly on the source model used in the fitting.
Thus, changes in the emission structure of the supernova during
its evolution could affect the model fitting estimates of the
source size, and hence the determination of the supernova
expansion, especially if a time-independent model were used.

The source structure remains circularly symmetric (with deviations
from circularity of less than 2\% over almost 4000 days). Using our 
data at 3.6 and 6\,cm, we can characterize the decelerated expansion of
SN\,1993J until day 1500 after explosion with a single expansion
parameter $m= 0.845\pm0.005~(R \propto t^{m})$. However, from that
day onwards the expansion appears different when observed at 6 and
18\,cm. At the latter wavelength, the expansion can indeed be 
characterized well by the same $m$ as before day 1500 (self-similar
expansion), while at the former wavelength the expansion appears
more decelerated and is characterized by another expansion
parameter, $m_{6}= 0.788\pm0.015$. From about day 1500 onwards,
the radio source size is progressively smaller at 6\,cm than at
18\,cm.

Our interpretation is that the expansion of the supernova is
self-similar and characterized by a single expansion parameter,
$m= 0.845\pm0.005$, which applies to all data before day 1500
after explosion and to only 18\,cm data all the time. In our
interpretation, the 6\,cm results would not represent the advance of
the supernova radius because at 6\,cm the size measured would be
systematically smaller than the source size. We can think of
several ways in which this could happen:

\begin{itemize}

\item Changing opacity affecting the emission at 6\,cm from the part of
the shell behind the supernova ejecta. Before about day 1500 after
explosion, the supernova ejecta would be opaque to the radiation
at all wavelengths but from then onwards the opacity at 6\,cm would
start to decrease, while remaining unchanged at 
18\,cm. This
hypothesis receives additional support from the evolution in the
total flux density at 6\,cm.

\item Radially decreasing magnetic field in the emitting region.
The consequences of the radially decreasing magnetic field would
come either from the limited sensitivity of the radio
interferometers or from synchrotron aging, or from both. In the
first case, emission from the outer regions detectable at 18\,cm
would fall below the detection limit at 6\,cm. In the second case,
the emission structure would be wavelength dependent.

\end{itemize}

Combining the two methods, i.e., CPM and model fitting, we determined
from 6\,cm data a fractional width of the radio shell of
$31(\pm2)$\%, and a degree of absorption of the radio emission
from the backside of the shell behind the ejecta of 80\%.
Unfortunately, we can determine from our data neither a possible
evolution in the absorption at 6\,cm nor the degree of absorption
at 18\,cm. For $80\%$ absorption at 18\,cm, we estimate a fractional
shell width at 18\,cm of $33.5(\pm1.7)~\%$. For $100\%$ absorption,
the fractional shell would be somewhat larger: $37.8(\pm1.3)$\%.
In both cases, the shell at 18\,cm is expected to be thicker than
at 6\,cm, given our interpretation of the different apparent
expansions at 6\,cm and 18\,cm for epochs beyond day 1500. These
findings differ from the Bartel et al. (\cite{Bartel2002}) results on the
details of the radio structure of SN\,1993J and its expansion.

For a distance of 3.63\,Mpc to SN\,1993J, comparison of our VLBI
results with optical spectral line velocities shows that the
deceleration is more pronounced in the radio than in the optical.
This difference in the deceleration might be due to a progressive
penetration of ejecta instabilities into the shocked circumstellar
medium, as also suggested by Bartel et al. (\cite{Bartel2007}).

\acknowledgements{
The National Radio Astronomy Observatory is a facility of the
National Science Foundation operated under cooperative agreement
by Associated Universities, Inc. The European VLBI Network is a
joint facility for European, Chinese, South African and other radio
astronomy institutes funded by their national research councils.
Partial support from Spanish grant AyA~2005-08561-C03-02 is
ackwnowledged. IMV is a fellow of the Alexander von 
Humboldt Foundation. 
KWW wishes to thank the Office of Naval Research Laboratories for 
the 6.1 funding supporting this research.}

\appendix

\begin{onecolumn}

\section{The Common Point Method}
\label{CPM-App}

Fitting a source model to the visibilities has some advantages: it
is possible to describe the information contained in many
visibilities with only a few parameters; and it is applied in
Fourier space, a natural choice for interferometric
measurements, thus avoiding the use of deconvolution algorithms
that might introduce artifacts in the source structure. However,
model fitting to the visibilities also has disadvantages. For
example, the fitted parameters depend on the details of the model
used. Also, if a model shape is kept fixed while source structure
changes take place, the fitting will introduce changing biases.
Additionally, there can also be strong coupling between the fitting
parameters, and the $\chi^2$ to be minimized may have lots of local
minima and/or a pathological behavior near the absolute minimum.

Measuring in the sky plane may sometimes (e.g., when the source is
very circularly symmetric) simplify matters: the measurement of
the source size is made directly on the source image. We describe
a new method for estimating the size of a circularly symmetric
source in the sky plane without assuming any a priori source
model. The algorithm is easy to use and rather insensitive to
changes in the internal structure of the observed source.

\subsection{Derivation}

We consider a circularly symmetric brightness distribution,
$M$, convolved by a Gaussian, $G$, of width $\sigma$. Let
$I(\sigma,x,y)$ be the resulting intensity distribution, where
$(x,y)$ are the arclengths of the relative right ascension and
declination. Then,

\begin{equation}
I(\sigma,x,y) = \int_{-\infty}^{\infty}\int_{-\infty}^{\infty}
M(x',y')G(x - x', y - y')dx' dy'
\end{equation}

\noindent where,

\begin{equation}
G(x,y) = \frac{1}{2\pi\sigma^2} \exp{\left ( - \frac{x^2 +
y^2}{2\sigma^2} \right )}
\end{equation}

If we express $I(\sigma,x,y)$ in polar coordinates and perform the
azimuthal average we have

\begin{equation}
Av(\sigma,r) = \exp{\left ( -\frac{r^2}{2\sigma^2} \right ) }
\int_{0}^{\infty} \frac{M(r')}{\sigma^2} \exp{\left (
-\frac{r'^{2}}{2\sigma^2} \right ) } \textrm{BesselI}_0 \left (
-\frac{r r'}{\sigma^2} \right ) r' dr', \label{Average}
\end{equation}

\noindent where $Av(\sigma,r)$ is the azimuthal average and
BesselI$_0$ is the modified Bessel function of the first kind. We
can consider Eq. \ref{Average} as the integral
transform of the brightness distribution, $M(r)$, in terms of the
following kernel function:

\begin{equation}
Kernel(\sigma,r,r') = \frac{r'}{\sigma^2} \exp{ \left ( -\frac{r^2
+ r'^2}{2\sigma^2} \right ) } \textrm{BesselI}_0 \left ( -\frac{r
r'}{\sigma^2} \right )  \label{Kernel}
\end{equation}

We can now compute the change in $Av(\sigma,r)$ for a small change
in $\sigma$. Performing a first-order Taylor expansion with
respect to $\delta\sigma$ (i.~e., the change in $\sigma$), we
obtain

\begin{equation}
\delta \left [ Av(\sigma,r) \right ] = \frac{\delta\sigma}{\sigma}
\int_{0}^{\infty} M(r') K(\sigma,r,r') dr' \label{Variation}
\end{equation}

\noindent with

\begin{equation}
K(\sigma,r,r') = 2 \frac{\exp{ \left ( -\frac{r^2 +
r'^2}{2\sigma^2} \right ) } r' dr'} {\sigma^2} \left ( \left
(\frac{r^2 + r'^2}{2\sigma^2} - 1 \right ) \textrm{BesselI}_0
\left ( \frac{r r'}{\sigma^2} \right ) - \frac{r r'}{\sigma^2}
\textrm{BesselI}_1 \left ( \frac{r r'}{\sigma^2} \right ) \right )
\label{KernelOfVariation}
\end{equation}

Thus, we can see from Eq. \ref{Variation} that if there is a value of $r$
(denoted $r_c$), directly related to $M$ and $\sigma$, such that

\begin{equation}
\int_{0}^{\infty} M(r') K(\sigma,r_c,r') = 0, \label{CPCondition}
\end{equation}

\noindent then the angular average $Av(\sigma,r)$ will not change
its value for small changes in the width of the convolving
Gaussian for $r=r_c$. We call {\em Common Point(s)} those points
of $Av(\sigma,r)$ that correspond to such value(s) of $r_c$. For
SN\,1993J, the values $r_c$ correspond to the abscissas $X_{ICP}$
and $X_{OCP}$ shown in Fig. 5 for observations from day 1889 after
the explosion.

For the purpose of discussion only, we assume that the profile
given by $M(r)$ has a clearly defined cutoff radius, $R$. Then, the
integral in Eq. \ref{Variation} will extend from $r' = 0$ to $r' = R$,
and $r_c$, multiplied by a factor $C$ that depends only on
$\sigma$ and $M(r)$, will equal $R$

\begin{equation}
r_c C(\sigma,M) = R \label{CP-wrt-R}
\label{A8}
\end{equation}

If we expand the brightness distribution, $M(r)$, self-similarly
by a factor $P$, then

\begin{equation}
R \rightarrow PR \Rightarrow M(r) \rightarrow M \left (
\frac{r}{P} \right ) \label{ExpansionP}
\end{equation}

\noindent and we arrive at a new expression for Eq. \ref{CPCondition}:

\begin{equation}
\int_{0}^{P R} M\left ( \frac{r'}{P} \right ) K \left ( \sigma,
\frac{r_c}{P},\frac{r'}{P} \right ) = 0 \label{CPCondition2}
\end{equation}

Given that both $r$ and $r'$ are scaled by $\sigma$ wherever they
appear in the kernel of Eq. \ref{KernelOfVariation}, the radial positions
of the Common Points associated with $M \left ( \frac{r}{P} \right
)$ and $P \sigma$ will be equal to $P$ times the radial positions
of the Common Points associated with $M(r)$ and $\sigma$. In other
words, given a self-similar expansion of a brightness
distribution, the Common Points will expand at the same rate as
the brightness distribution, provided that the Gaussians used in
the convolutions are also scaled with the source size. In such
cases, the relationship between $r_c$ and $R$ would be given by
Eq. \ref{CP-wrt-R}, where $C$ would only depend, for the whole
expansion, on the profile of the brightness distribution, $M(r)$
and the (constant) ratio between the source radius and the
convolving beam. We computed, using $\sigma = 0.5 R$, the
values of $C$ for a set of possible source distributions, always
finding values of $C$ near unity (see Table \ref{table-bias} for six
examples of $C$, which we call {\em bias} and label as
$R/X_{OCP}$).

Thus, the only condition that must be satisfied for using the
Common Point to determine the expansion of SN\,1993J is that, for
each epoch, the convolving Gaussian beam must be equal to the
supernova size multiplied by a given factor, which must be the
same for all epochs. The Gaussian beam for each epoch can be found
in an iterative way, given that the Common Points are very stable
to changes in the convolving beam. We provide the details of this
iterative process in the next appendix section.

\subsection{Application}

We convolve a map corresponding to a shell-like source of
unknown outer radius $R$ with a Gaussian of width $\sigma$. Let us
also convolve the same map with Gaussians of different widths
given by $\sigma_i = H_i \sigma$, where the constants $H_i$ are
all near unity (we use 0.8, 0.9, 1.1, and 1.2).

If we now compute the azimuthal average of each of these maps and
superimpose the results, we find that the profiles cross in
narrow regions of radial values\footnote{As can be seen in Fig. 5,
there are two crossing regions, one at the inner edge and the
other at the outer edge of the profile. For determining source
sizes, we are primarily interested in the outer crossing region.}.
The mean value of the radial positions of the crossing points will
be an estimate of the radial position of the Common Point
described above. Let this estimate be $r_{c,1}$.

Now convolve the initial map with a new set of Gaussians, starting
with a new $\sigma = F r_{c,1}$, where $F$ is a chosen constant
(in practice, we choose $F = 0.5$) and with $\sigma_i = H_i
\sigma$, where the constants $H_i$ are the same as in the previous
iteration. We now find a new value for the radial position of the
Common Point: $r_{c,2}$. We can further iterate this procedure to
obtain $r_{c,3}, r_{c,4}, ...$ The procedure will soon converge to
a value $r_{c,f}$. By construction, this value is the common point
associated with the brightness profile of the shell and a
convolving beam of width equal to $F$ times the (unknown) source
radius. Thus, this iterative process gives the radial position of
the Common Point associated with a Gaussian beam that scales
always by the same factor $F$ with respect to the source radius,
regardless of the size of the source.

The Common Point Method works very well with synthetic data. Because of 
the practical insensitivity of the computed Common Point to the
sizes of the Gaussian widths (within reasonable limits, of
course), virtually independent of the radial profile of the source
structure, the restriction of using $H_i$ close to unity can be
lifted. Azimuthal averages of VLBI maps of SN\,1993J, constructed
using Gaussian beamwidths ranging from $\sigma$ to $1.5\sigma$,
will, when superimposed, cross at radial positions that differ by
a very small amount (typically, $\sim$1$\mu$as) (see Fig. \ref{fig42}).

This makes the Common Point Method a robust way of measuring the
outer radius of a circularly symmetric, and self-similarly
expanding, source. It converges in only a few iterations, largely
independent of the starting value of $\sigma$.

As a further test of the method, we generated synthetic data 
at several observing frequencies, with different shell sizes, and 
for different dynamic ranges. In all cases, the CPM estimates 
were excellent compared to the true values. We also studied 
the CPM bias for different (non-homogeneous emission) shell 
structures, for example, a parabolic-shaped shell (zero at extremes and maximum 
at center), a double mini-shell (shell with emission only in $R_{in}$ 
and $R_{out}$), and a shell with a linear radial decay. In all 
cases, the CPM bias (i.e., the value of $C$ in Eq. \ref{A8}) is close 
to 1, and is also invariant in a self-similar expansion.

For a real map, we determine $r_{c,f}$. Since we know from our
testing that $C$ is close to unity, we can very reliably assign
the value $r_{c,f}$ to $R$. If the expansion is exactly
self-similar, the bias $C$ remains constant at all epochs. The
accuracy of the source size determination for all epochs is thus
translated to an enhanced accuracy in the determination of the
expansion curve of the source. Even for slight departures from
self-similar expansion, the method remains more accurate than
model fitting, according to our testing. If the expansion departs
from self-similar, the bias $C$ will slightly change, but will
change by much less (usually by a factor of 2, depending on how
the evolution of the brightness profile $M(r)$ differs from
self-similarity) than in the model fitting. For example, a change
in a shell with no ejecta opacity from a fractional shell width of
0.25 to 0.35 (i.e., a change of 40\% in the fractional shell
width) translates into a bias change of only 1.5\% for $X_{OCP}$
(see Table \ref{table-bias}). However, the same change in the fractional
shell width would translate into a bias change of around 3\%
(depending on the source size; the larger the source, the smaller
the bias change) if we apply model fitting to the visibilities.

\subsection{Source size uncertainties}

The CPM itself does not have a clearly defined way of estimating 
the source size uncertainty. The remarkable property that all profiles
obtained with different beams cross at $X_{OCP}$ does not have a 2
dimensional equivalent, except for the ideal case of perfect
circular symmetry. We assume that the uncertainty in the
determination of the source size is related to the departure from
circularity of the source. Hence, we assign an uncertainty as described 
below.

We call ``MAP1'' the map reconstructed with a beam size of half the
radius of SN\,1993J (as Fig. \ref{fig41}(a)). We then
reconstruct a map, ``MAP2'', using a beam size 10\% larger (as
Fig. \ref{fig41}(b)) than that used to reconstruct MAP1 (the
method is almost insensitive to the choice of this percentage). A
subtraction of MAP2 from MAP1 will yield ``MAP3'' (as in Fig.
\ref{fig41}(c)). To estimate the departure from circularity of the
source, we measure from MAP3 the scatter in the radial values of
the outer zeroth contour level, $\zeta$, along 800 directions
equally distributed in azimuth (the number of directions is
arbitrary, and the estimate is insensitive to this choice provided
the number is higher than a few hundred). To minimize a possible
bias in this measurement between epochs, we reconstruct the map
with a pixel size proportional to the beam size, which in turn is
proportional to the source size, as mentioned earlier. The standard
deviation assigned to the source size measurement is thus $\zeta$.

\begin{figure}[!t]
\centering
\includegraphics[width=6cm]{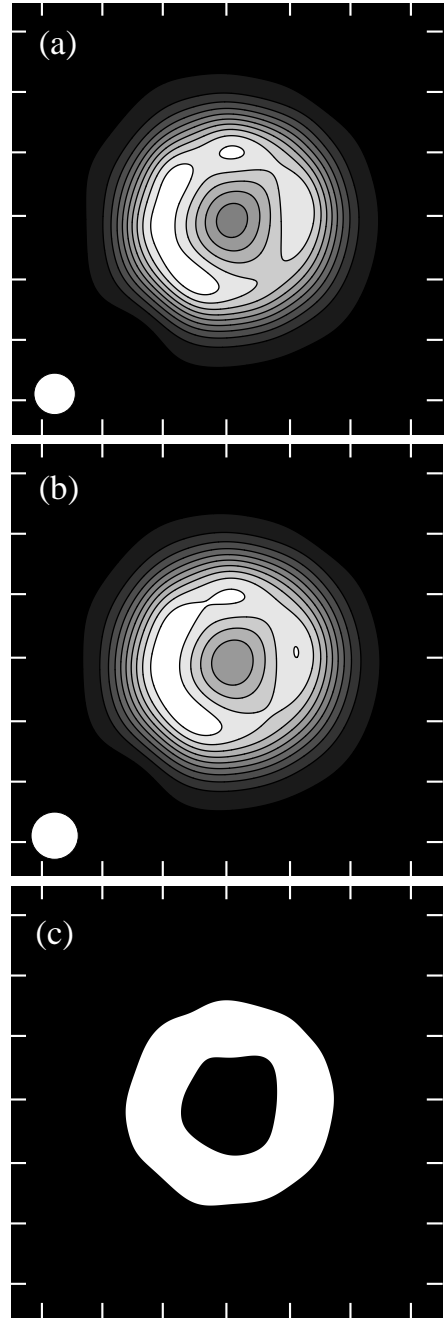}
\caption{6~cm contour maps from day 1889 reconstructed using (a) a
1.74 mas beam; (b) a 1.92 mas beam. The ten contours shown are
linearly spaced from 2.56 to -0.28 mJy beam$^{-1}$ for (a) and
from 2.91 to -0.32 mJy beam$^{-1}$ for (b). The bottom image, (c)
shows the result of the subtraction of (b) from (a); residuals
outside the source have been removed in maps (a) and (b) prior to
subtraction. Image (c) was then digitized in flux to 1 bit
(positive is white, negative is black). The transition contour of
interest in image (c) is the outer contour. Tick marks are in
units of 2mas.} \label{fig41}
\end{figure}

\section{Visibility modelfitting}
\label{Modelfit-App}

The radial profile of the projection in the sky of the emission
from a spherical shell with absorption in its inner side (i.e.,
from intervening ejecta) is given by the expression

%\begin{equation}
%I(A,R,W,a,r) =
%\begin{cases}
%A\,(\sqrt{R^2-r^2} - \sqrt{(R\,W)^2 - r^2})\,(2 - a)/2  &  r < R\,W \\
%A\,(\sqrt{R^2-r^2})  &  r \geq R\,W
%\end{cases}
%\label{ModI}
%\end{equation}

\begin{equation}
I(A,R,W,a,r) = \left\{\begin{array}{rl}
A\,(\sqrt{R^2-r^2} - \sqrt{(R\,W)^2 - r^2})\,(2 - a)/2 & \textrm{if}~~~r < R\,W \\
      \\
A\,(\sqrt{R^2-r^2}) & \textrm{if}~~~r \geq R\,W
\end{array} \right.
\label{ModI}
\end{equation}

\noindent where $r$ is the radial coordinate, $R$ is the source
radius, $W$ the fractional radius of the inner surface of the
shell, $a$ is the degree of absorption (0 for no absorption; 1
for total absorption), and $A$ is a scaling factor related to the
total shell flux density. The real part of the azimuthal average
of the visibilities is fit to  $H$, the Hankel transform of $I$,
which is equal to the azimuthal average of the Fourier transform
of a circularly symmetric source with a profile given by $I$.
Thus, the model used in the fitting is

\begin{equation}
H(q) =
\int_0^{\infty}{I(A,R,W,a,r)\,BesselI_0(2\,\pi\,q\,r)\,r\,dr}
\label{ModV}
\end{equation}

\noindent where $q = \sqrt{u^2+v^2}$ is the distance in Fourier
space. The parameters $A$, $R$, $W$, and $a$ are fitted but not
necessarily all everytime. We can fix some parameters and fit the
others, as explained in several sections of this paper. As said in
Sect. \ref{Fourier-Sec}, we apply a radial binning to the
visibilities during the azimuthal average and downweight the long
baselines with a taper prior to the model fit. We use 300 bins,
which cover the first 5 amplitude lobes (see Fig. \ref{fig31}
for a schematic representation of these lobes). For epochs where
the supernova was not large enough, the width of the bins is scaled
according to the source size, to obtain a similar radial coverage
of each averaged visibility (in units of the sizes of the lobes)
for all epochs. In Fig. \ref{ModFitFig}, we show an example of a
model fit using $R$ and $A$ as fitting parameters and fixing $a$ to
1 and $W$ to 0.7 (i.e., the fractional shell width to 0.3).

\begin{figure}[!h]
\centering
\includegraphics[width=10cm]{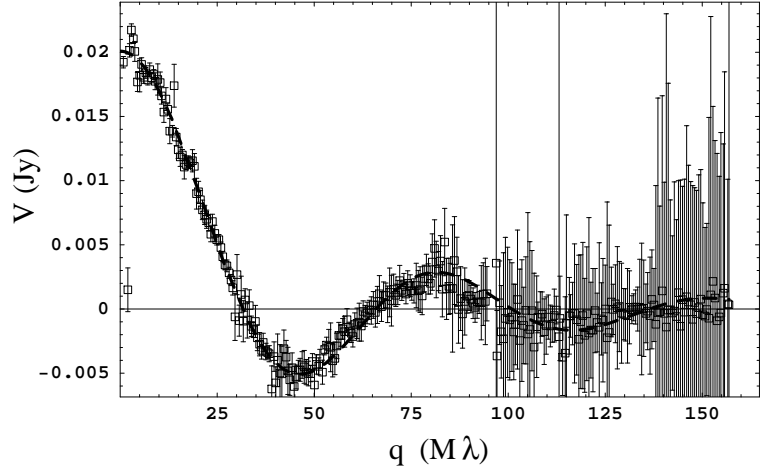}
\caption{Model fit of a spherical shell (of fractional width 0.3
and total absorption, $a$ = 1) to the visibilities, observed on
day 1889 after explosion, which have been azimuthally averaged as
described in this appendix. The fitting parameters are the source
size $R$ and the parameter related to the flux density, $A$. The
fitted model is shown as a dashed line. Notice the increase in the
error bars for the longest baselines, due to the taper applied.}
\label{ModFitFig}
\end{figure}

\section{Emission structure due to a synchrotron-aged electron population}
\label{Ageing-App}

The spectrum of the synchrotron emission of an electron with
energy $E$ in a magnetic field $B$ is (Pacholczyk, \cite{Pacholczyk1970})

\begin{equation}
I_{\nu}(B,E) \propto
B\cdot\textrm{F}\left(\sqrt{\frac{3}{2}}\frac{\nu}{c_{1}Â·BÂ·E^{2}}
\right)
\end{equation}

\noindent where $c_{1} = 6.27 \times 10^{18}$ (cgs units) and
we assume that $B^{2}=\frac{3}{2}B_{\perp}^{2}$,
$B_{\perp}$ being the magnetic field component orthogonal to the
electron velocity.

For large values of $x$, the function F$(x)$ behaves as follows:

\begin{equation}
\textrm{F}(x) \propto \sqrt{x}\cdot\exp{(-x)}
\end{equation}

\noindent Thus, for small enough values of $B\,E^{2}$, the ratio of
emission at frequencies $\nu_{1}>\nu_{2}$ is

\begin{equation}
\frac{I_{\nu_{1}}(E)}{I_{\nu_{2}}(E)} =
\sqrt{\frac{\nu_{1}}{\nu_{2}}}
\cdot\exp\left(\sqrt{\frac{3}{2}}\frac{(\nu_{2}-\nu_{1})}{c_{1}Â·BÂ·E^{2}}\right)
\label{intensityratio}
\end{equation}

\noindent As shown, this intensity ratio depends on the strength of the magnetic
field. If we now consider the spectrum resulting from
integrating over all energies using an electron energy
distribution of the canonical type $N \propto E^{-p}$, the
intensities $I_{\nu}$ at different frequencies will be such that
the dependence of the intensity ratio on the magnetic field will
vanish since the resulting spectrum will depend separately on $B$
and $\nu$. However, this separation would not hold were the
distribution lower than the canonical at high energies as in
synchrotron aging. In this case, after integration over all
electron energies, the intensity ratio depends on the magnetic
field. Thus, a radially changing magnetic field would translate
into a radially changing ratio of emission intensities at
different frequencies.

For day 3200 after explosion, Chandra et al. (\cite{Chandra2004})
inferred a spectral index $\alpha$ that is steeper at higher
frequencies than at lower frequencies, i.e.,

\begin{equation}
\alpha = \left\{ \begin{array}{ll}
0.51 & \nu \leq 4\textrm{~GHz}; \\
1.13 & \nu > 4\textrm{~GHz}.\end{array} \right. %}
\label{chandrapop}
\end{equation}
%\alpha = \twochoices{0.51,&&\nu \leq 4 \times 10^{9}} {1.13,&&\nu
%> 4 \times 10^{9}}

\noindent We now compute the electron distribution $N(E)$ that reproduces
the spectrum measured by Chandra et al. (\cite{Chandra2004}) and obeys Eq.
(1). The power index, $p$, of the energy distribution is related
to the spectral index, $\alpha$, of the emission spectrum by
the expression $p = 1+2\alpha$ (Pacholczyk \cite{Pacholczyk1970}). We can now
integrate over all energies to obtain the contribution of all the
electrons to the emission intensity at each distance $D$ (distance
from the contact discontinuity) and radiofrequency $\nu$:

\begin{equation}
S_{\nu}(D)= \int_{E_{min}}^{\infty}{I_{\nu}(B(D),E) N(E) dE}
\end{equation}

\noindent The normalized profile of this emission intensity is shown in
Fig. \ref{fig-theor-intens} for 6 and 18\,cm. For a given value
of $D$, the emission at the longer wavelength is always higher
than the emission at the shorter wavelength. In other words, for a
given intensity level, the emission reaches further out at the
longer wavelength.

\end{onecolumn}

\end{document}